\documentclass[fleqn,usenatbib]{mnras}

\usepackage{newtxtext,newtxmath}

\usepackage[T1]{fontenc}
\usepackage{ae,aecompl}
\usepackage{pdflscape}
\usepackage{afterpage}

\usepackage{graphicx}	

\usepackage{amsmath}	
\usepackage{amssymb}	
\usepackage{subfig}
\usepackage{booktabs}
\usepackage{afterpage}
\usepackage{float}
\usepackage{tablefootnote}
\usepackage{hyperref}
\usepackage{threeparttable}
\usepackage[pagewise,columnwise]{lineno}

\newcommand{\mysim}{\mathord{\sim}}
\newcommand{\myapprox}{\mathord{\approx}}
\newcommand{\logp}{\log P}

\title[Cepheid amplitudes]{Reassessing the Constraints from SH0ES Extragalactic Cepheid Amplitudes on Systematic Blending Bias}

\author[A. Sharon et al.]{
	Amir Sharon$^{1}$\thanks{E-mail: amir.sharon@weizmann.ac.il}, Doron Kushnir$^{1}$, Wenlong Yuan$^{2}$, Lucas Macri$^{3}$ and Adam Riess$^{2}$
	\\
	$^{1}$Department of Particle Physics \& Astrophysics, Weizmann Institute of Science, Rehovot 76100, Israel\\
	$^{2}$Department of Physics and Astronomy, Johns Hopkins University, Baltimore, MD 21218, USA\\
	$^{3}$NSF's National Optical Infrared Astronomy Research Laboratory, Tucson, AZ 85726, USA\\
}

\date{Accepted XXX. Received YYY; in original form ZZZ}

\pubyear{2021}

\begin{document}
\label{firstpage}
\pagerange{\pageref{firstpage}--\pageref{lastpage}}
\maketitle

\begin{abstract}
The SH0ES collaboration Hubble constant determination is in a $\mysim5\sigma$ difference with the \textit{Planck} value, known as the Hubble tension. The accuracy of the Hubble constant measured with extragalactic Cepheids depends on robust stellar-crowding background estimation. Riess et al. (R20) compared light-curve amplitudes of extragalactic and MW Cepheids to constrain an unaccounted systematic blending bias, $\gamma=-0.029\pm0.037\,\rm{mag}$, which cannot explain the required, $\gamma=0.24\pm0.05\,\rm{mag}$, to resolve the Hubble tension. Further checks by Riess et al. demonstrate that a possible blending is not likely related to the size of the crowding correction. We repeat the R20 analysis, with the following main differences: (1) we limit the extragalactic and MW Cepheids comparison to periods $P\lesssim50\,\rm{d}$, since the number of MW Cepheids with longer periods is minimal; (2) we use publicly available data to recalibrate amplitude ratios of MW Cepheids in standard passbands; (3) we remeasure the amplitudes of Cepheids in NGC 5584 and NGC 4258 in two \textit{Hubble Space Telescope} filters ($F555W$ and $F350LP$) to improve the empirical constraint on their amplitude ratio $A^{555}/A^{350}$. We show that the filter transformations introduce an $\myapprox0.04\,\rm{mag}$ uncertainty in determining $\gamma$, not included by R20. While our final estimate, $\gamma=0.013\pm0.057\,\rm{mag}$, is consistent with the value derived by R20 and is consistent with no bias, the error is somewhat larger, and the best-fitting value is shifted by $\myapprox0.04\,\rm{mag}$ and closer to zero.  Future observations, especially with JWST, would allow better calibration of $\gamma$.
\end{abstract}




\section{introduction}
\label{sec:introduction}

The latest determination of the Hubble constant by the SH0ES collaboration \citep[][hereafter R22]{Riess2022}, $H_0=73.04\pm1.04\,\rm{km}\,\rm{s}^{-1}\,\rm{Mpc}^{-1}$, is in a $\mysim5\sigma$ difference with the \textit{Planck} value \citep{Planck2020}, $H_0=67.4\pm0.5\,\rm{km}\,\rm{s}^{-1}\,\rm{Mpc}^{-1}$, known as the Hubble tension. The difference between the Cepheid and Type Ia supernovae-based SH0ES measurement and the cosmic microwave background temperature and polarization anisotropies \textit{Planck} measurement has led to numerous suggestions for extensions of the standard Lambda cold dark matter ($\Lambda$CDM) cosmology model \citep[see][for a review]{DiValentino2021}. The SH0ES absolute distance scale is based on the period-luminosity relation of Cepheids \citep[$P-L$ relation;][]{Leavitt1912} measured in the \textit{Hubble Space Telescope} $F160W$ filter (similar to the near-infrared $H$ band). The Cepheids reside in $37$ Type Ia supernovae host galaxies and other anchor galaxies with an absolute distance measurement. The Hubble tension can be expressed as $\mysim0.1-0.2\,\textrm{mag}$ difference in the magnitudes of SH0ES Cepheids \citep{Riess2019,Efstathiou2020}, in the sense that the SH0ES Cepheids (in M31 and further away) are brighter than the $\Lambda$CDM prediction.


The accuracy of the Hubble constant measured with extragalactic Cepheids depends on robust photometry and background estimation in the presence of stellar crowding. The SH0ES collaboration performs artificial Cepheid tests and derives a crowding correction, $\Delta m_H$, which is added to the photometry of each Cepheid (i.e., reducing the brightness of the Cepheid). \citet[][hereafter R20]{Riess2020} pointed out that crowding by unresolved sources at Cepheid sites reduces the fractional amplitudes of their light curves. This is because the crowding adds a constant sky background flux that compresses the relative flux amplitude variations of a Cepheid. R20 compared the \textit{HST} $F160W$ amplitudes of over 200 Cepheid amplitudes in three hosts (hereafter \textit{faraway galaxies}) and in the anchor galaxy NGC 4258 to the observed amplitudes in the Milky Way (MW). This comparison allowed them to constrain a possible systematic bias in the determination of the crowding correction, $\gamma=-0.029\pm0.037\,\rm{mag}$\footnote{Note a typo in R20, with reported $\gamma=0.029\pm0.037\,\rm{mag}$.}, which cannot explain the required systematic error to resolve the Hubble tension. Note that the results of R20 suggests that both the calculated crowding correction, which estimated the chance superposition of Cepheids on crowded backgrounds, is accurate and that light from stars physically associated with Cepheids \citep[with the prime candidates being wide binaries and open clusters;][]{Anderson2018} is small. In other words, R20 constrained the total systematic blending bias to be $\gamma=-0.029\pm0.037\,\rm{mag}$.

In this paper, we repeat the analysis of R20 with a careful study of each step required for the comparison of the extragalactic amplitudes to the MW amplitudes. The main differences between our analysis and the analysis of R20 are:
\begin{itemize}
\item We impose the period limit $\logp\equiv \log_{10}(P\,[\rm{d}])<1.72$ for the comparison (the period range of the R20 extragalactic Cepheids is $1<\logp<2$), since the amplitudes for longer period Cepheids cannot be reliably determined for the MW (as the number of such MW Cepheids is minimal, see Appendix~\ref{sec:properties}). We obtain similar results by removing the period cut but adding increased uncertainties to the MW relations at long periods.
\item We use public available data to recalibrate amplitudes ratios of MW Cepheids in standard bands along with their associate uncertainties. We show that a calibration of the required filter transformations from Cepheid observations introduces an $\myapprox0.04\,\rm{mag}$ uncertainty in the determination of $\gamma$, not included by R20. We show that available Cepheids templates are not accurate enough to reduce this error. Our transformation between two \textit{HST} filters ($F555W$ and $F350LP$; $A^{555}/A^{350}$) is different from the transformation used by R20. We show that the transformation used by R20 did not optimally weight the data, and we calibrate a new transformation based on updated amplitude measurements.
\end{itemize}

Our final estimate for a possible blending bias is $\gamma=0.013\pm0.057\,\rm{mag}$. While the obtained $\gamma$ is consistent with the value derived by R20 and is consistent with no bias, the error is somewhat larger, and the best-fitting value is shifted by $\myapprox0.04\,\rm{mag}$. To be clear, the measurement of $\gamma$ is not a component in the direct determination of $H_0$ from the distance ladder nor is it a quantity measured in other experiments such as by Planck. Rather it is a parameter used to construct a specific null test of the hypothesis of unrecognized Cepheid crowding.  The fact that our result is consistent with zero means we can only say the null test regarding this hypothesis is passed, rather than using it to provide a new value of $H_0$ or of the Tension.  (Section~\ref{sec:discussion}). 

The method of R20 to compare the extragalactic amplitudes to the MW amplitudes is described in Section~\ref{sec:R20 method}. In Section~\ref{sec:HST amplitude} we calibrate the required \textit{HST}filters transformation and in Appendix~\ref{sec:Ground-HST transformation} we calibrate the required ground-HST filter transformations. In Section~\ref{sec:amp_extra} we repeat the analysis of R20 using our methods. We discuss some caveats of our analysis and the implications of our results in Section~\ref{sec:discussion}. 

We independently recalibrate MW Cepheids amplitude ratios by constructing a galactic Cepheid catalogue from publicly available photometry (Appendix~\ref{sec:catalog}). We employ Gaussian processes (GP) interpolations on the phase-folded light curves to determine the mean magnitudes and amplitudes in different bands. 

We follow the convention that a single Cepheid magnitude $x$ is the magnitude of intensity mean, $x=\langle x \rangle$, and colours $(x-y)$ stand for $\langle x \rangle-\langle y \rangle$. All fits in this paper includes global $2.7\sigma$ clipping. In order to decide on the optimal polynomial order for the fitting, we normalized the errors to obtain a reduced $\chi^2$ of $1$, and we inspect the difference $\Delta\chi^2$ obtained with a higher-by-one order polynomial.  

\section{The method of R20}
\label{sec:R20 method}

R20 compared between the amplitudes of extragalactic Cepheids and MW Cepheids to constrain a possible systematic blending bias, $\gamma$. Specifically, R20 minimized
\begin{equation}\label{eq:gamma}
\chi^{2}\left(\gamma\right)=\sum_{i=1}^{n}\left(\frac{A^{160}_i}{A^{350}_i}-\frac{A^{160,\rm{MW}}}{A^{350,\rm{MW}}}10^{-0.4\left(\Delta m_{i,H}-\Delta m_{i,V}+\gamma\right)}\right)^2\sigma_{i}^{-2},
\end{equation}
where the summation is over all extragalactic Cepheids in the sample (see R20 for a derivation of Equation~\eqref{eq:gamma}). $A^{160}_{i}$ and $A^{350}_{i}$ are the observed amplitudes\footnote{The amplitude is defined here as the magnitude difference between the minimum and the maximum of the light curve.} of the extragalactic Cepheids in $F160W$ and the white filter $F350LP$, respectively. These amplitudes were evaluated by fitting the \citet{Yoachim2009} light curve templates to the photometric data that is usually noisy and sparse, see details in Section~\ref{sec:HST amplitude}. The term   $A^{160,\rm{MW}}/A^{350,\rm{MW}}$ is the calibrated transformation between these amplitudes (that depends on the period of the Cepheid), which is based on accurate amplitude measurements of MW Cepheids in standard passbands and on a transformation to the \textit{HST} filters, see below. The factor $10^{-0.4(\Delta m_{i,H}-\Delta m_{i,V})}$ is the expected reduction in the amplitude ratio because of crowding, which also depends on the (small) crowding correction in the $F350LP$ filter, $\Delta m_{i,V}$, and $\sigma_i$ is the relevant error of the expression. The motivation to study the $A^{160}/A^{350}$ amplitude ratio instead of the near-infrared (NIR) amplitude is the reduction in the observed scatter around the MW relation ($\myapprox0.05$, see Section~\ref{sec:amp_rat_HV}, compared with $\myapprox0.1\,\rm{mag}$ for the NIR amplitude). The Cepheids in the sample have $1<\logp<2$ with $A^{160}\sim0.2\,\rm{mag}$ (measured with an accuracy of $\mysim0.1\rm{mag}$), compared with $A^{160,\rm{MW}}$ in the range of $0.2-0.5\,\rm{mag}$ for the same period range. The crowding corrections for the Cepheids in the sample are mostly $\Delta m_{H}\lesssim0.6\,\rm{mag}$ with a smaller fraction of Cepheid found in regions with higher surface brightness (up to $\Delta m_{H}\approx2\,\rm{mag}$) than the limit typically used to measure $H_0$.

The transformation of the MW relation, observed in $H$ and $V$ bands, to the \textit{HST} filters is performed in R20 with
\begin{equation}\label{eq:ratio transformation}
\frac{A^{160,\rm{MW}}}{A^{350,\rm{MW}}}=\frac{A^{H,\rm{MW}}}{A^{V,\rm{MW}}}\frac{A^{160}}{A^{H}}\frac{A^{V}}{A^{555}}\frac{A^{555}}{A^{350}},
\end{equation}
where $A^{555}$ is the amplitude in the $F555W$ filter (similar to the $V$ band). The ratios $A^{160}/A^{H}$ and $A^{V}/A^{555}$ can be determined by comparing ground-based observations to \textit{HST} observations \citep[see][and references therein]{Riess2021a}. The ratio $A^{555}/A^{350}$ can be determined from \textit{HST} observations of extragalactic Cepheids. R20 used $A^{160}/A^{H}=1.015$, $A^{555}/A^{V}=1.04$\footnote{The relation $A^{V}/A^{555}=1.04$ in R20 is a typo.} and the transformation from \citet[hereafter H16]{Hoffmann16} for the $A^{555}/A^{350}$ ratio. By a minimization of Equation~\eqref{eq:gamma}, R20 obtained $\gamma=-0.029\pm0.037\,\rm{mag}$, which cannot explain the required systematic error to resolve the Hubble tension. 

Here, we repeat the analysis of R20 with a careful study of each step required for the comparison of the extragalactic amplitudes to the MW amplitudes. The values of $A^{160}_{i}$, $A^{350}_{i}$, $\Delta m_{i,H}$, $\Delta m_{i,V}$, and $\sigma_{i}$ are taken from Table 3 of R20\footnote{Note that the NGC 4258 Cepheid amplitudes were measured with the $F555W$ filter, so the transformation of H16 between $A^{350}$ and $A^{555}$ was used to derive the values in Table 3 of R20. Also, the provided $\sigma_{i}$ (and their properties, described below Equation (14) of R20) were multiplied by $A^{350}_{i}$, so one should divide the provided error by $A^{350}_{i}$, which is typically smaller than 1, to be used in Equation~\eqref{eq:gamma}.}. The transformation $A^{H,\rm{MW}}/A^{V,\rm{MW}}$ is rederived in Section~\ref{sec:amp_rat_HV}, including the uncertainty of this transformation. While the rederived transformation is similar to the result of R20, the uncertainty has a significant contribution to the final uncertainty of $\gamma$, which was not considered by R20. The $A^{555}/A^{350}$ transfomration is rederived in Section~\ref{sec:HST amplitude}. Our transformation is different from the transformation used by R20. Finally, the $A^{160}/A^{H}$ and $A^{V}/A^{555}$ transformations are rederived in Appendix~\ref{sec:Ground-HST transformation}. Although our method is different from the method of R20 for these two transformations, we find similar results and the uncertainty of the transformations has a small contribution to the final uncertainty of $\gamma$. A summary of the sources and derivations of the terms in Equations~\eqref{eq:gamma} and~\eqref{eq:ratio transformation} is provided in Table~\ref{tab:sources}.

\begin{table*}
    \centering
    \caption{Summary of sources and derivations of the terms in Equations~\eqref{eq:gamma} and~\eqref{eq:ratio transformation}.}
    \label{tab:sources}
    \begin{threeparttable}
    \begin{tabular}{cccccl} 
    \hline
     Term & Source & relation to R20 & comments \\ \hline
       $A^{160}_i$  & R20 & -  & -  \\
       $A^{350}_i$  & R20 & -  & -  \\
       $\Delta m_{i,H}$  & R20 & - & - \\
       $\Delta m_{i,V}$  & R20 & - & - \\
       $\sigma_{i}$      & R20 & - & - \\
       $A^{H,\rm{MW}}/A^{V,\rm{MW}}$  & Section~\ref{sec:amp_rat_HV} & similar & incl. significant uncertainty (not considered by R20) \\
       $A^{555}/A^{350}$  & Section~\ref{sec:HST amplitude} & different &  incl. small uncertainty (not considered by R20)\\
       $A^{160}/A^{H}$  & Appendix~\ref{sec:Ground-HST transformation} & similar & incl. small uncertainty (not considered by R20)  \\
       $A^{V}/A^{555}$  & Appendix~\ref{sec:Ground-HST transformation} & similar & incl. small uncertainty (not considered by R20)  \\
    
    \end{tabular}
\end{threeparttable}
\end{table*}

\section{The MW $A^{H}/A^{V}$ ratio}
\label{sec:amp_rat_HV}

In this section, we use our catalogueue (see Appendix~\ref{sec:catalog}) to derive the $A^{H}/A^{V}$ amplitude ratio of the MW Cepheids with $1<\logp<1.72$. Amplitude ratios of other bands that are used to estimate the ground-HST filter transformations in Appendix~\ref{sec:Ground-HST transformation} are presented in Appendix~\ref{sec:amp_rat}.

\begin{figure}
	\includegraphics[width=\columnwidth]{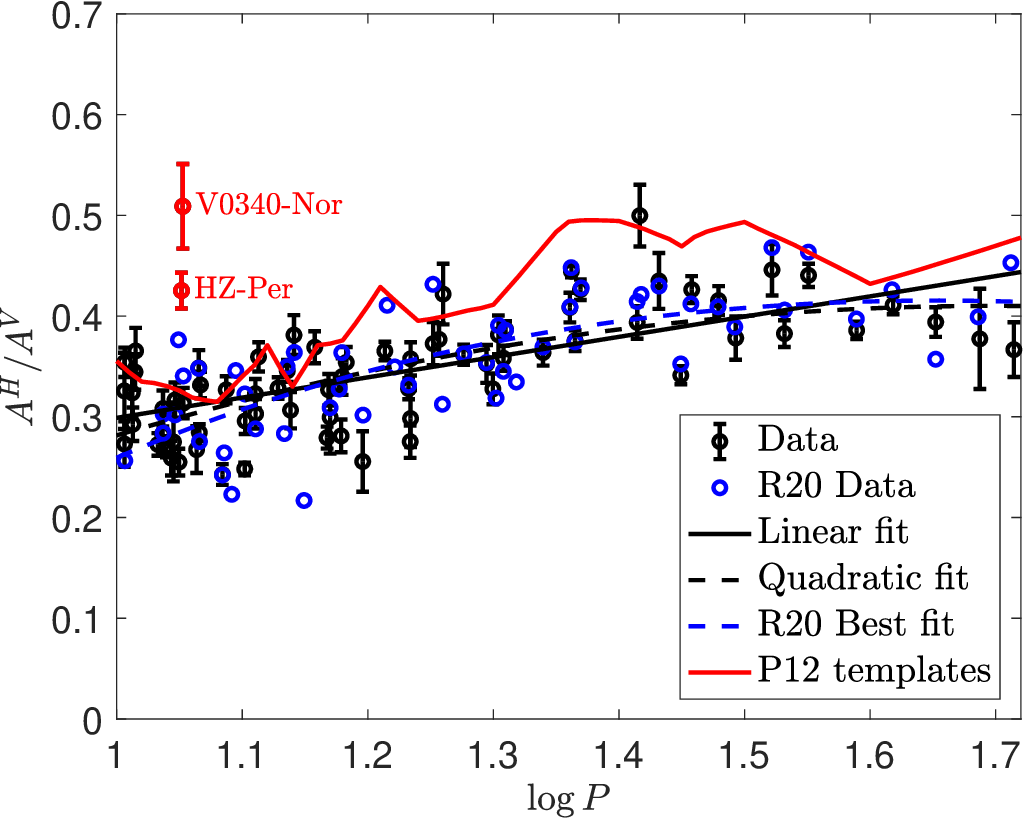}
	\caption{The distribution of $A^{H}/A^{V}$ as a function of the period. The observations (black symbols) are well fitted with a first-order and second-order polynomials in $\logp$ (solid and dashed black lines, respectively). The P12 templates (red line) overpredict the fitted functions by $\lesssim25\%$. The data points from Table 1 of R20 and the best-fitting second-order polynomial derived in R20 are plotted in blue. The best-fitting of R20 and the one derived here are similar.}
	\label{fig:AHoAV}
\end{figure}

The $A^{H}/A^{V}$ ratio of different MW Cepehids as a function of period is presented in Figure~\ref{fig:AHoAV} as black symbols. Note that in most cases, each amplitude is derived from high signal-to-noise ratio photometric data that is available in a large number of epochs. We fit for the transformation a linear function (solid black line). We find in this case $\chi^2_{\nu}\approx9.8$ for $75$ Cepheids after the removal of the outlier V0340-Nor, suggesting an intrinsic scatter of $\myapprox0.037$. The results of the fit following the addition of the calibrated intrinsic scatter is $(0.20\pm0.03)(\logp-1)+(0.30\pm0.01)$\footnote{the off-diagonal term in the covariance matrix of the linear fit is $\myapprox-1.55\times10^{-4}$, which is required for the analysis in Sections~\ref{sec:HST amplitude}-\ref{sec:amp_extra}.}. We find a small improvement for fitting with a quadratic function, $\Delta\chi^2\approx2.9$, i.e. less than $2\sigma$\footnote{Incuding the longest period Cepheid with $A^{H}/A^{V}$ measurement, S-Vul with $\logp=1.84$, to the sample changes $\Delta\chi^2$ to $\myapprox4.4$ between the quadratic and the linear fit, indicating a larger but still insignificant ($\myapprox2\sigma$) improvement.}. Nevertheless, we also use a quadratic function (as used by R20) to check the sensitivity of our results. We find for the quadratic fit (dashed black line) $\chi^2_{\nu}\approx8.3$ for $74$ Cepheids after the removal of the outliers V0340-Nor and HZ-Per, suggesting an intrinsic scatter of $\myapprox0.034$. The results of the fit following the addition of the calibrated intrinsic scatter is $(-0.26\pm0.13)(\logp-1)^2+(0.37\pm0.08)(\logp-1)+(0.28\pm0.01)$. The result of the \citet[][hereafter P12]{Pejcha2012} templates are presented as well (red line) and it overpredict the fitted functions by $\lesssim25\%$\footnote{The deviation of the P12 templates are probably related to the fact that the data of \citet{Monson2011} were not included in the P12 fitting, while it dominates our $H$-band catalogueue (O. Pejcha, private communication). The deviation also suggests that the accuracy of the P12 templates for the amplitude in a wide filter, such as $F350LP$, is limited, see Section~\ref{sec:HST amplitude}.}. We also plot the data points from Table 1 of R20\footnote{Note that VZ-Pup has a double entry in Table 1 of R20 and that the period of SV-Vul should be $\myapprox45\,\rm{d}$ and not as stated there ($P=14.10\,\rm{d}$).} and the best-fitting second-order polynomial derived in R20 (blue)\footnote{The best-fitting for the $A^{H}/A^{V}$ ratio is derived from the best-fitting for the $A^{160}/A^{350}$ ratio, given in R20, multiply by the filter transformation functions, as given in R20.}. The best-fitting of R20 and the quadratic fit derived here are similar. 

We reproduced the known result that the $\gtrsim0.1\,\rm{mag}$ scatter seen in single-band amplitudes can be significantly reduced by considering amplitude ratios between different bands \citep[][and references therein]{Klagyivik2009}. This was the motivation of R20 to study the ratio $A^{H}/A^{V}$ 

\section{The $A^{555}/A^{350}$ ratio}
\label{sec:HST amplitude}

In this section, we discuss the amplitude transformation $A^{555}/A^{350}$, which is required for the comparison in Section~\ref{sec:amp_extra} (see Equation~\eqref{eq:ratio transformation}), and significantly affects the estimation of $\gamma$. Unlike the situation with ground filter amplitudes, where high signal to noise ratio photometric data is available in a large number of epochs, the calibration of \textit{HST} filter amplitudes is less certain and involves template fitting. As we demonstrate below, there are different calibrations of the $A^{555}/A^{350}$ ratio that deviate significantly from each other. Before we discuss the actual observations, we provide some intuition for the expected ratio and the predictions of available templates.

Assume that the Cepheid at the time of maximum (minimum) light is a blackbody with a temperature $T_{h}$ ($T_{l}$), with typical values $6000-7000\,\rm{K}$ ($4400-5000\,\rm{K}$) \citep[see, e.g., Figure 3 of][hereafter J21]{Javanmardi2021}. We can use the well observed relation $A^{I}/A^{V}\approx0.6$ to determine a relation between $T_{h}$ and $T_{l}$ (regardless of the radii of the Cepheid at extremum light)\footnote{Note that the filter $F350LP$ contains a significant overlap with the $V$ and $I$ bands, see Figure 2 of H16.}. From the relation between $T_{h}$ and $T_{l}$ we find $A^{555}/A^{350}\approx1.07-1.12$. In principle, well calibrated templates can provide a more accurate estimate for this ratio. However, the results from the P12 templates, $A^{555}/A^{350}\approx1.04$\footnote{Since no observations with \textit{HST} filters were used to calibrate the P12 templates, the predictions of the P12 templates for these filters depends on theoretical atmospheric models. We use the values $\beta=5.42,5.62,1.7$ for the $F350LP$, $F555W$ and $F160W$ filters, kindly provided to us by Ond\v{r}ej Pejcha (see equation (3) of P12 for details). We discuss a method to improve the P12 templates predictions for \textit{HST} filters in Section~\ref{sec:discussion}.}, and from the templates used by J21, $A^{555}/A^{350}\approx1.17$, deviate by more than $10\%$. This deviation is probably related to the large wavelength range ($\mysim0.3-1\,\mu\rm{m}$, see Figure 2 of H16), which the white filter $F350LP$ spans, that is challenging to describe accurately (see, e.g., the $\mysim25\%$ deviation of the $A^{H}/A^{V}$ between the P12 templates and observations in Section~\ref{sec:amp_rat_HV}), and to the less precise prediction of the P12 templates for \textit{HST} filters (see the $\mysim5\%$ deviation of the $A^{555}/A^{V}$ between the P12 templates and our estimate in Appendix~\ref{sec:Ground-HST transformation}). 

In what follows, we discuss various analyses of the data from $F555W$ and $F350LP$ observations, to estimate the $A^{555}/A^{350}$ transformation. In Section~\ref{sec:empirical} we discuss an empirical calibration based on a sample of Cepheids in NGC 5584. In Section~\ref{sec:other} we discuss other, less robust, methods. We summerize our findings in Section~\ref{sec:sum rat}. 

\subsection{NGC 5584 empirical calibration}
\label{sec:empirical}

\begin{figure*}
	\includegraphics[width=0.7\textwidth]{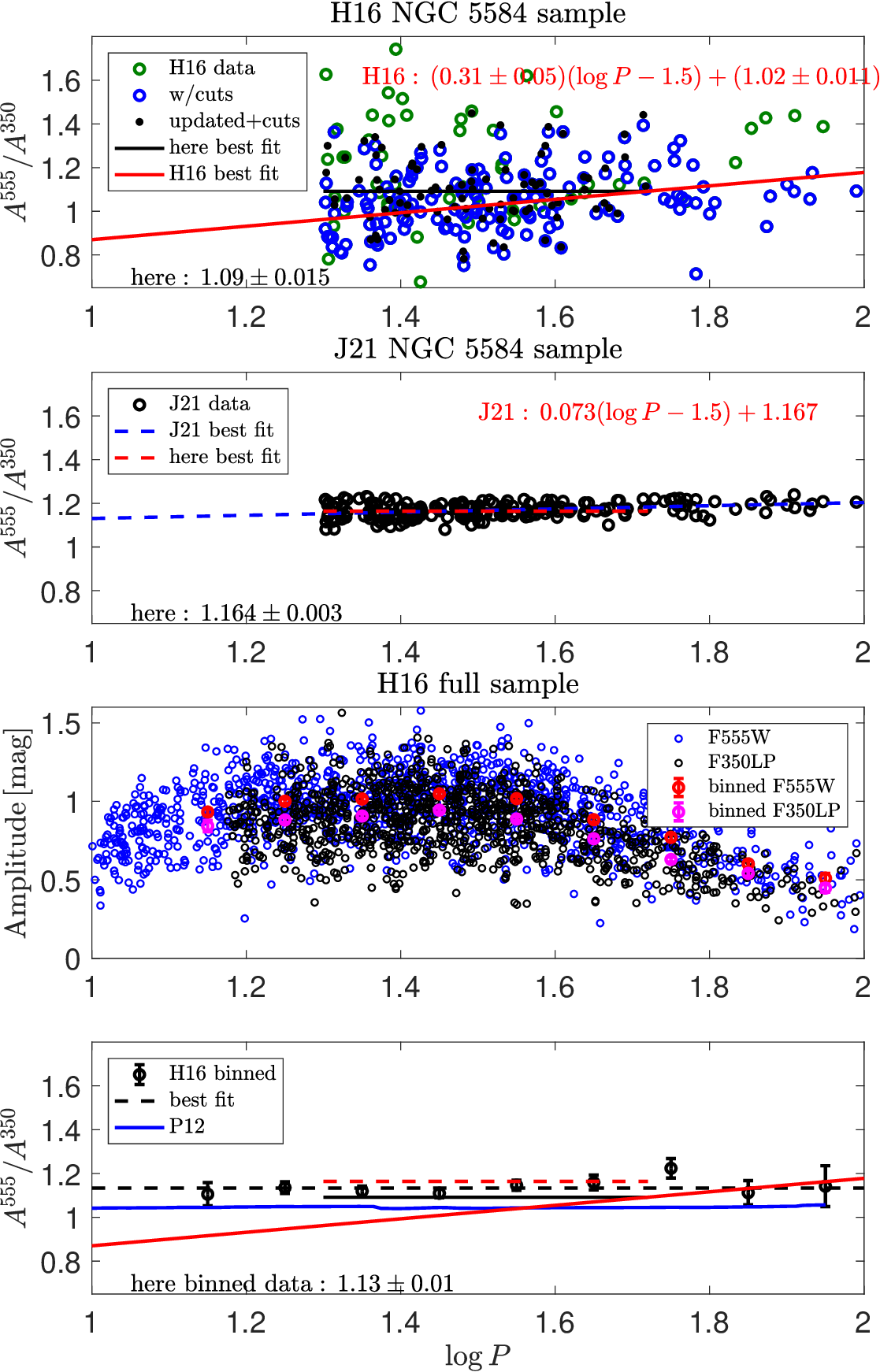}
	\caption{The $A^{555}/A^{350}$ transformation. Top panel: the $A^{555}/A^{350}$ ratio for Cepheids in NGC 5584 with $A^{555}$ and $A^{350}$ determinations for each Cepheid as a function of period. Green symbols: all $199$ Cepheids above a period cut. Blue symbols: only Cepheids that passed the additional H16 cut. Black symbols: the recalibrated amplitudes that passed our cuts. Black line: our best-fitting to the Cepheids that passed our cuts. Red line: the H16 transformation. The H16 transformation satisfies $A^{555}/A^{350}<1$ for $\logp\lesssim1.4$ and $A^{555}/A^{350}=0.87$ for $\logp=1$, while it is expected that $A^{555}/A^{350}\gtrsim1$. Second panel: The NGC 5584 analysis of J21 (black symbols). J21 found a very small $A^{555}/A^{350}$ scatter, $\myapprox0.015$, and they fitted the data with $0.073(\logp-1.5)+1.167$ (blue dashed line). The obtained $A^{555}/A^{350}$ scatter is significantly smaller than the scatter in any MW amplitude ratio, suggesting that the J21 fit to the NGC 5584 Cepheids is artificially constrained by their MW templates. We limit the data to $\logp<1.72$  and we obtain $A^{555}/A^{350}=1.164\pm0.003$ by fitting a constant value to the data (red dashed line). Third panel: the full sample of H16, which includes $1325$ Cepheids with $A^{555}$ values (blue) and $1035$ Cepheids with $A^{350}$ values (black). We bin the data in the range $1.1\le\logp\le2$ with a bin size of $0.1$ and find the mean and scatter in each bin for each filter (red and magenta symbols with error bars). Bottom panel: The means ratio as a function of period (black symbols). Dashed black line: the best-fitting to the binned data, $A^{555}/A^{350}=1.13\pm0.01$. For reference, our best-fitting for the NGC 5584 sample (solid black line), the H16 fit (solid red line), our fit to the J21 data (dashed red line) and the P12 templates (blue line) are shown as well. Except for the H16 fit, which is unreliable, all estimates suggest a constant ratio for $A^{555}/A^{350}$ (or a very weak period dependence), with the range of $1.04-1.16$. The best estimate that we have for this ratio is our best-fitting for the NGC 5584 sample (solid black line), see text.}
	\label{fig:350Vs555}
\end{figure*}

R20 used a transformation that was derived in H16 from a sample of Cepheids in NGC 5584 with $A^{555}$ and $A^{350}$ determinations for each Cepheid (we remind the reader that the individual amplitudes were evaluated by fitting the \citet{Yoachim2009} light curve templates to the photometric data that is usually noisy and sparse). The derived transformation by H16 was $A^{555}/A^{350}=(0.308\pm0.052)(\logp-1.5)+(1.024\pm0.011)$\footnote{Note a typo in Table 2 of H16.}, with a scatter of $0.134$, presented as red line in the top panel of Figure~\ref{fig:350Vs555}. This transformation satisfies $A^{555}/A^{350}<1$ for $\logp\lesssim1.4$ and $A^{555}/A^{350}=0.87$ for $\logp=1$, which significantly deviates from our expectation above. These amplitudes are publicly available only for $199$ Cepheids above a period cut, and they are plotted in green symbols. For the fit in H16, additional cuts were imposed on the data\footnote{$1<A^{555}/A^{814}<2.2$, $0.55< A^{555}/A^{350}<1.4$, $0.35< A^{814}/A^{350}<0.975$.}, and the data that passed these cuts are plotted in blue symbols. To reconstruct the transformation of H16, we applied the same cuts for the publicly available Cepheids, and obtained a best-fitting of $A^{555}/A^{350}=(0.24\pm0.07)(\logp-1.5)+(1.05\pm0.01)$ (with a scatter of $\myapprox0.13$ for $148$ Cepheids after clipping; the linear fit performs significantly better than a constant ratio), which is similar to the H16 fit. The small inconsistency of our fit and the H16 fit could be explained with the few additional Cepheids of H16. While both the fit preformed here and the fit of H16 suggest a slope that is significant by more than $3\sigma$, there are a few issues with this procedure. 

First, it is evident that the errors of the amplitude ratios are dominated by the light-curve fitting to a noisy photometric data in a small number of epochs (H16 did not provide error bars), as the intrinsic scatter should be roughly bounded by the scatter of the $A^{I}/A^{V}$ relation, $\myapprox0.06$ (see Appendix~\ref{sec:amp_rat}), while the obtained scatter is larger by a factor of $\myapprox2$. The implication is that different error-bars should be assigned to each Cepheid, as one cannot assume that the observed scatter is dominated by the intrinsic scatter. The procedure of H16 is to assume a constant error bar for all data, which is equivalent to assuming a good fit. As a result, one cannot get an independent goodness-of-fit probability, making the significance of the slope statistically meaningless. Moreover, the periods of the NGC 5584 sample are $\logp>1.3$, and the extrapolation to shorter periods of the positive slope fit amplifies the deviation between the fit and the expectations. Indeed, $40$ light curves in NGC 4258 with  $0.7< \logp<1.4$ (mean $0.95$) measured by \citet{Yuan2022} in both $F555W$ and $F350LP$ yield a mean $A^{555}/A^{350}$ value of $1.068 \pm 0.022$, strengthening the case of a constant ratio with period. A second issue with the determination of $A^{555}/A^{350}$ from the NGC 5584 sample is the use of cuts. While the motivation to use cuts in order to remove unreliable results (e..g, blending with a nearby source that changes the amplitude ratio) is well justified, the exact choice of the cuts affect the obtained $A^{555}/A^{350}$. For example, in the case that no cuts are employed on the data, we obtain a best-fitting of $A^{555}/A^{350}=1.09\pm0.01$ (with a scatter of $\myapprox0.17$ for $196$ Cepheids after clipping; we find no significant improvement for fitting with a linear function). This transformation is significantly larger than the H16 transformation for $\logp\lesssim1.5$, see the black line that represents a similar transformation. This large deviation is driven by the tendency of the H16 cuts to remove observations with large value of $A^{555}/A^{350}$, evident by comparing the green to the blue symbols in the figure. 

The amplitudes and amplitude ratios measured in H16 used a coarse grid $(0.01)$ to identify the best-fit amplitudes (of the photometry to the \citet{Yoachim2009} templates) with no mapping of the $\chi^2$ space to measure individual uncertainties, thus errors were assumed constant. Here, we reevaluate the amplitudes for the Cepheids in NGC 5584 using a higher resolution grid sampling of $0.001$ (benefiting from faster CPUs than available in 2015), and employ a mapping of the $\chi^2$ space to determine individual uncertainties. \footnote{The revised amplitudes are publicly available \url{https://drive.google.com/drive/folders/1pCWp0_QARVE6EzsDSI5bMOaKdvIRHV6D?usp=sharing}.}. Using the revised data and similar quality cuts in H16 (to reduce the impact of blending)\footnote{$0.35<A^{814}/A^{555}<0.85$, $0.55< A^{555}/A^{350}<1.45$, $0.3< A^{814}/A^{350}<0.9$.} and the full period range yields a constant amplitude ratio $A^{555}/A^{350}=1.074 \pm 0.011$, and weak evidence of a trend with $\logp$ ($0.052\pm0.035$), in good agreement with the results of \citet{Yuan2022}. We further limit the data to $\logp<1.72$ and additionally require a small crowding bias, $|\Delta m_I|<0.05\,\rm{mag}$, as derived by J21 for the $F814W$ filter\footnote{We thank B. Javanmardi for kindly providing us with the crowding biases calculated in J21.} (black symbols). We obtain $A^{555}/A^{350}=1.09\pm0.015$ by fitting a constant value to the data (black line; we find $\chi^2_{\nu}\approx0.90$ for $88$ Cepheids and a scatter of $\myapprox0.15$; we find no significant improvement for fitting with a linear function).

In appendix~\ref{sec:simulation} we perform simulations of the process of measuring amplitude ratios, $A^{555}/A^{350}$, in a distant galaxy like NGC 5584, as was done in H16. We find a small, $\mysim0.015$, overestimate of the amplitude ratio from measured data. We have not corrected the empirical estimate of the mean amplitude ratio for this bias but we make a note of it here.

One issue with our estimate form above is related to the use of light curve templates for the light curve fitting. We used the \citet{Yoachim2009} $V$ band templates for fitting both the $F555W$ and the $F350LP$ photometry, where the amplitude in each band is allowed to change freely, providing an empirical estimate for the amplitude ratio. However, while the use of the $V$ band templates to estimate $A^{555}$ is justified due to the similarity of the $V$ band and $F555W$ filters, it is not clear that the use of the $V$ band template to estimate $A^{350}$ does not introduce any bias. The $F350LP$ light curve shape has not been measured accurately, and, as we demonstrated above, it is challenging for available templates to accurately describe the behaviour of such a wide filter. We note that we found no change in the amplitude ratio by substituting the $B$-band \citet{Yoachim2009} template for the $V$-band to fit $F350LP$ light curves, so this ratio does not appear particularly sensitive to the shape of the template within reason. We discuss methods to improve this situation in Section~\ref{sec:discussion}. We adopt $A^{555}/A^{350}=1.09\pm0.015$ as our best estimate at the moment, but keeping in mind the caveat with this method, we describe other methods to estimate $A^{555}/A^{350}$ in the following Section. 

\subsection{Other methods}
\label{sec:other}

The same NGC 5584 observations were also analyzed by J21, in which an independent light-curve modelling approach has been implemented (see details below). They found a very small $A^{555}/A^{350}$ scatter, $\myapprox0.015$, and they fit the data with $0.073(\logp-1.5)+1.167$, see the second panel of Figure~\ref{fig:350Vs555}. We limit the data to $\logp<1.72$  and we obtain $A^{555}/A^{350}=1.164\pm0.003$ by fitting a constant value to the data (dashed red line; we find a scatter of $\myapprox0.03$ for $170$ Cepheids; we find no significant improvement for fitting with a linear function). The $A^{555}/A^{350}$ values obtained in this method are higher by $\myapprox7\%$ from the estimate in the previous section, which is based on the SH0ES collaboration light-curve modelling. As we explain below, the J21 estimate is less reliable, since it heavily relies on their MW templates, and the accuracy of their templates is expected to be lower than $\myapprox10\%$.   

The light-curve modelling approach of J21 for a given Cepheid includes a simultaneous fit of all bands to their MW templates. These templates already include some pre-determined $A^{555}/A^{350}$ amplitude ratio (we emphasize that the $F350LP$ light curve shape has never been measured for any MW Cepheid), and their fitting process do not allow each light curve to independently determine its own amplitude and thus to measure the amplitude ratio directly from the data. This situation is evident from the fit results of J21. First, their fitted line passes directly through the results of their MW templates, and second, the obtained $A^{555}/A^{350}$ scatter is significantly smaller than the scatter in any MW amplitude ratio (see Appendix~\ref{sec:amp_rat}), suggesting that the J21 fit to the NGC 5584 Cepheids is artificially constrained by their MW templates. While the J21 MW templates are not publicly available, their prediction for $A^{814}/A^{555}$ (see their equation 3), can be compared with minimal manipulation to the measured $A^{I}/A^{V}$. We show in Appendix~\ref{sec:amp_rat} that they overpredict the observed ratio by $\gtrsim10\%$. This result suggests that the ability of J21 MW templates to predict $A^{555}/A^{350}$, which include a challenging modelling of a wide filter, is limited by (at least) $\myapprox10\%$. 

Another estimate for $A^{555}/A^{350}$ can be obtained with the full sample of H16, which includes $1325$ Cepheids with $A^{555}$ values and $1035$ Cepheids with $A^{350}$ values. We emphasize that most of the data is obtained for different Cepheids (except for the $199$ Cepheids in NGC 5584 with both $A^{555}$ and $A^{350}$ values), which limit the robustness of the results from this sample, as we explain below. The sample is plotted in the third panel of Figure~\ref{fig:350Vs555}. We bin the data in the range $1.1\le\logp\le2$ with a bin size of $0.1$. We find the mean and scatter in each bin for each filter (red and magenta symbols with error bars). We find that the means of $A^{555}$ are consistently larger than the means of $A^{350}$. The amplitude distributions in each period bin are given in Appendix~\ref{sec:histogram}, where it is evident that the entire $A^{555}$ distribution is shifted from the $A^{350}$ distribution to higher amplitudes. We next plot the means ratio as a function of the period (bottom panel of Figure~\ref{fig:350Vs555}), for which we can assign reliable errors. We fit the data with $A^{555}/A^{350}=1.13\pm0.01$ (dashed black line; we find no significant improvement for fitting with a linear function). This method can introduce a bias to the calibrated $A^{555}/A^{350}$ ratio, since the distribution of amplitudes in each bin is determined by the intrinsic amplitude distribution and by the observational error, which neither is accurately constrained for $F555W$ and for $F350LP$. While additional study is required to calibrate this bias, we apply various cuts to the full H16 data (ignoring M101 and NGC 4258 and/or using only Cepheids included in \citet[][hereafter R16]{Riess2016}), and we do not find a significant effect on the results. 

The bottom panel of Figure~\ref{fig:350Vs555} summarizes the different estimates. Except for the H16 fit, which is unreliable, all estimates suggest a constant ratio for $A^{555}/A^{350}$ (or a very weak period dependence), with the range of $1.04-1.16$. The best estimate that we have for this ratio is based on the updated measurements of the H16 amplitudes in NGC 5584, $A^{555}/A^{350}=1.09\pm0.015$, which is used as our preferred value in what follows (hereafter \textit{empirical}). As we explained above, this estimate is not free from caveats. We also demonstrate the sensitivity of our results by considering $A^{555}/A^{350}=1.15$, which represents the high-end range of estimates (hereafter \textit{speculative}). We emphasize that this high value is less reliable than our preferred value, and it is only considered for the purpose of demonstrating the sensitivity of our results to  $A^{555}/A^{350}$ and to motivate additional observations that will improve the accuracy of the $A^{555}/A^{350}$ calibration, discussed in Section~\ref{sec:discussion}. 

\subsection{Summary}
\label{sec:sum rat}

\begin{figure}
	\includegraphics[width=\columnwidth]{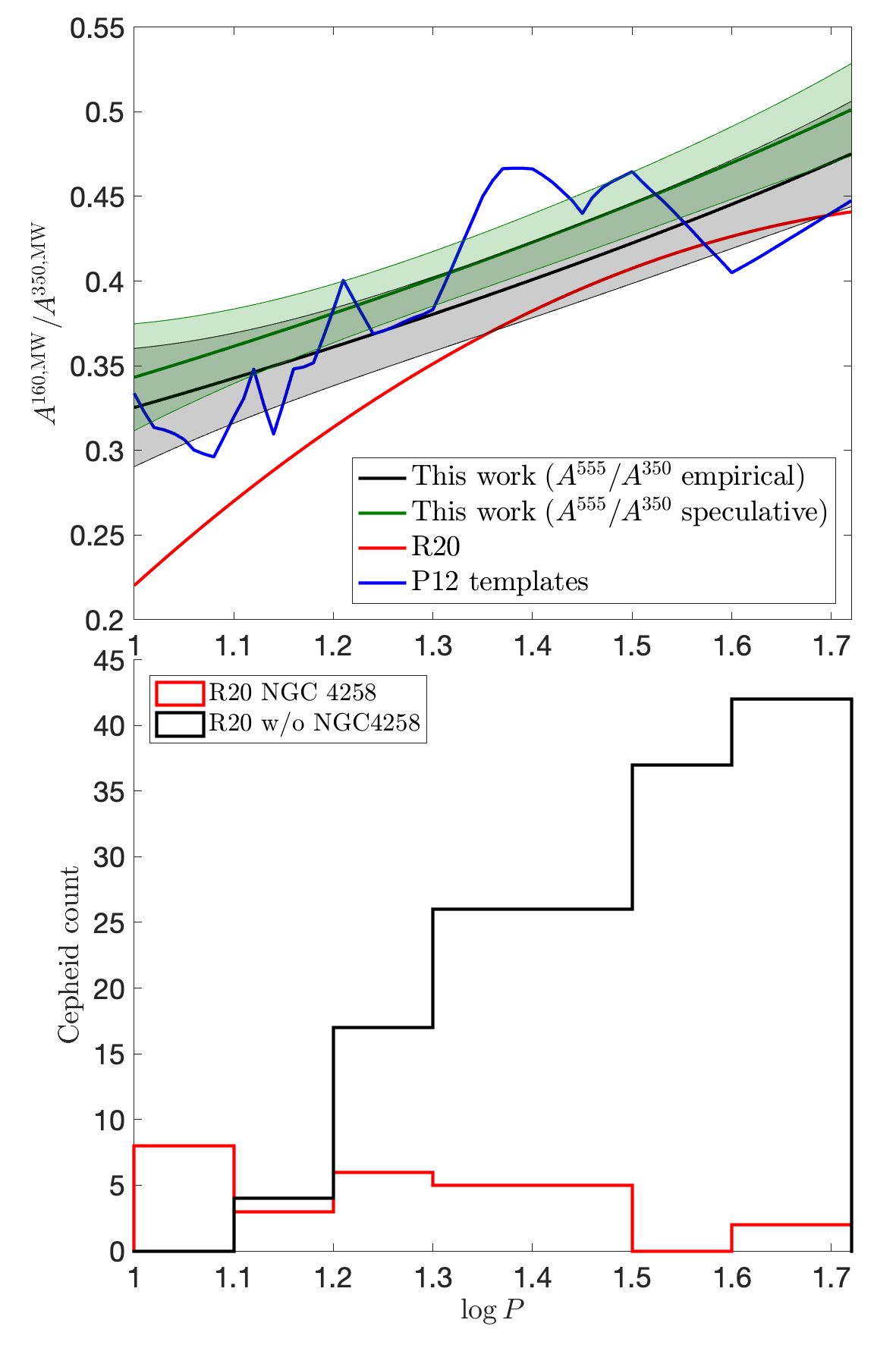}
	\caption{Top panel: a comparison of our derived $A^{160,\rm{MW}}/A^{350,\rm{MW}}$ (based on the empirical $A^{555}/A^{350}$ in black and based on the speculative $A^{555}/A^{350}$ in green) to the relation used by R20 (red line) and to the prediction of the P12 templates (blue line), as a function of $\logp$. Our estimation uses the terms (see Equation~\eqref{eq:ratio transformation}) $A^{H,\rm{MW}}/A^{V,\rm{MW}}=(0.20\pm0.03)(\logp-1)+(0.30\pm0.01)$ (see Appendix~\ref{sec:amp_rat}), $A^{160}/A^{H}$ and $A^{555}/A^{V}$ from Appendix~\ref{sec:Ground-HST transformation} (that are similar to the ratios of R20) and $A^{555}/A^{350}$ from this section. The black and green shaded areas represent the (systematic) uncertainty of the transformations, not considered by R20. Our derived relation is different from the relation used by R20, mostly because of the different $A^{555}/A^{350}$ transformation. Our derived relation agrees fairly well with the P12 templates prediction, but this is a coincidence, as there are significant deviations in some terms of Equation~\eqref{eq:ratio transformation} that cancel out.  Bottom panel: the R20 extragalactic sample distribution of periods with $\logp$ bin widths of $0.1$ (the last bin is between 1.6 and 1.72). Red: Cepheids in NGC 4258 that were measured with the $F555W$ filter, not requiring the $A^{555}/A^{350}$ transformation to compare with the MW. Black: Cepheids in the faraway galaxies that were measured with the $F350LP$ filter. The largest deviation between our results and R20 (at $\logp\lesssim1.2$) is effecting only a small number of Cepheids.}
	\label{fig:160Vs350}
\end{figure}

We conclude this section with the comparison in Figure~\ref{fig:160Vs350} of our derived $A^{160,\rm{MW}}/A^{350,\rm{MW}}$ (based on the empirical $A^{555}/A^{350}$ in black and based on the speculative $A^{555}/A^{350}$ in green) to the relation used by R20 (red line) and to the prediction of the P12 templates (blue line). Our estimation uses the terms (see Equation~\eqref{eq:ratio transformation}) $A^{H,\rm{MW}}/A^{V,\rm{MW}}=(0.20\pm0.03)(\logp-1)+(0.30\pm0.01)$ (see Section~\ref{sec:amp_rat_HV}), $A^{160}/A^{H}$ and $A^{555}/A^{V}$ from Appendix~\ref{sec:Ground-HST transformation} (that are similar to the ratios of R20) and $A^{555}/A^{350}$ from this section. The black and green shaded areas represent the (systematic) uncertainty of the transformations, not considered by R20. As can be seen in the figure, our derived relation is somewhat different from the relation used by R20, mostly because of the different $A^{555}/A^{350}$ transformation, as discussed in this section. Our derived relation agrees fairly well with the P12 templates prediction, but this is a coincidence, as there are significant deviations in some terms of Equation~\eqref{eq:ratio transformation} that cancel out. Also presented in the figure is the R20 extragalactic sample distribution of periods with $\logp$ bin widths of $0.1$ (the last bin is between 1.6 and 1.72). Cepheids in NGC 4258 that were measured with the $F555W$ filter, not requiring the $A^{555}/A^{350}$ transformation to compare with the MW, are presented in red. Cepheids in the faraway galaxies that were measured with the $F350LP$ filter are presented in black. The largest deviation between our results and R20 (at $\logp\lesssim1.2$) is effecting only a small number of Cepheids.

\section{Constraining a blending bias}
\label{sec:amp_extra}

In this section, we repeat the analysis of R20 that compares the $A^{160}/A^{350}$ amplitude ratios of extragalactic Cepheids to the $A^{H}/A^{V}$ amplitude ratios of MW Cepheids to constrain a possible systematic blending bias, $\gamma$, and the sensitivity of its value to various modifications proposed in this study. As in R20, this is done by minimizing Equation~\eqref{eq:gamma}\footnote{We hereafter assume that the errors are normally distributed.}. The results for various variants are presented in Table \ref{tab:gamma_values}.

\begin{figure}
	\includegraphics[width=\columnwidth]{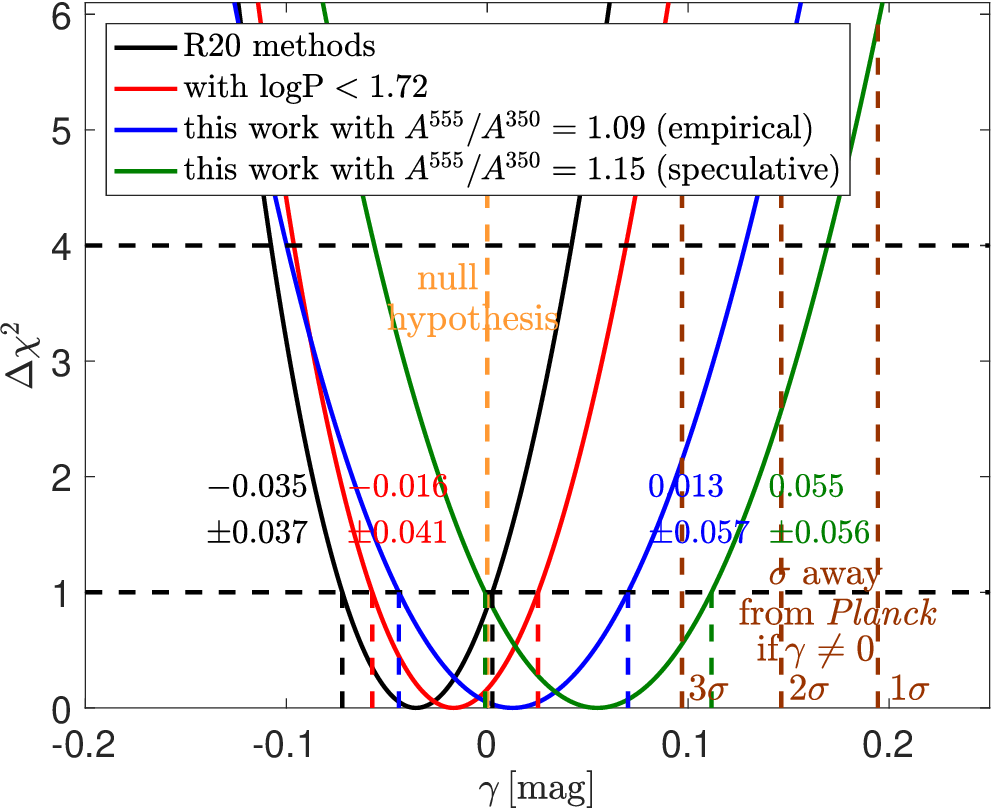}
	\caption{Constrains on a possible systematic blending bias, $\gamma$ with $\chi^2$ tests (Equation~\eqref{eq:gamma}). Black solid line: using the methods of R20. Red line: limiting the sample to Cepheids with $\logp<1.72$, as the MW relation cannot be determined reliably for larger periods (see Section~\ref{sec:properties}). Blue line: Additionally using $A^{160}/A^{H}$ and $A^{555}/A^{V}$ from Appendix~\ref{sec:Ground-HST transformation}, the expression for $A^{H,\rm{MW}}/A^{V,\rm{MW}}$ found in Appendix~\ref{sec:amp_rat}, and the empirical $A^{555}/A^{350}=1.09$ (see Section~\ref{sec:HST amplitude}) instead of the H16 relation. We find $\gamma=0.013\pm0.057\,\rm{mag}$ (including the transformations uncertainty). Green line: same as the blue line, but using the speculative $A^{555}/A^{350}=1.15$ (see Section~\ref{sec:HST amplitude}) instead of the H16 relation. We find $\gamma=0.055\pm0.056\,\rm{mag}$ (including the transformations uncertainty). Dashed brown lines: the distance (in $\sigma$s) from the \textit{Planck} results, in the case that gamma would be measured with high accuracy (see Section~\ref{sec:discussion} for details). In order to remove the Hubble tension, a value of $\gamma=0.24\pm0.05\,\rm{mag}$ is required. The $\gamma$ tests are consistent with $\gamma=0$ (orange line; the null hypothesis), and so we have not detected any evidence of a bias.}
	\label{fig:gamma}
\end{figure}

We first attempt to reproduce the analysis in R20, i.e., we do not apply a period cut and we use the MW relation derived in R20. We find $\gamma=-0.035\pm0.037\,\rm{mag}$ (variant 1, black line in Figure~\ref{fig:gamma}), in a good agreement with the value $\gamma=-0.029\pm0.037\,\rm{mag}$ obtained by R20. We next limit the sample to Cepheids with $\logp<1.72$, as the MW relation cannot be determined reliably for larger periods (see Section~\ref{sec:properties}). We find a $\myapprox0.02\,\rm{mag}$ increase $\gamma=-0.016\pm0.041\,\rm{mag}$ (variant 2, red line). We next use the expression for $A^{H,\rm{MW}}/A^{V,\rm{MW}}$ that was calibrated in Section~\ref{sec:amp_rat_HV} and the expressions for the $A^{160}/A^{H}$ and $A^{555}/A^{V}$ transformations from Appendix~\ref{sec:Ground-HST transformation}. These are similar to the ratios used by R20 and do not have a large effect. We additionally use the empirical $A^{555}/A^{350}=1.09$ (see Section~\ref{sec:HST amplitude}) instead of the H16 relation, and we find a $\myapprox0.03\,\rm{mag}$ increase, $\gamma=0.012\pm0.041\,\rm{mag}$ (variant 3). Not limiting the Cepheid periods to $\logp=1.72$ would lead to a smaller change, as the H16 relation predicts $A^{555}/A^{350}>1.09$ for $\logp>1.72$. Including the transformations uncertainty, see below, leads to our final result, $\gamma=0.013\pm0.057\,\rm{mag}$ (variant 4, blue line). Using the speculative $A^{555}/A^{350}=1.15$ (see Section~\ref{sec:HST amplitude}) instead of the H16 relation, we find a $\myapprox0.07\,\rm{mag}$ increase, $\gamma=0.054\pm0.041\,\rm{mag}$ (variant 5). Including the transformations uncertainty, see below, leads to our final result in this case, $\gamma=0.055\pm0.056\,\rm{mag}$ (variant 6, green line). The above results are consistent with $\gamma=0$ and so we have not detected any evidence of a bias.

\begin{table*}
    \centering
    \caption{Variants of the fit for a possible blending bias, $\gamma$. The R20 result is $\gamma=-0.029\pm0.037\,\rm{mag}$.}
    \label{tab:gamma_values}
    \begin{threeparttable}
    \begin{tabular}{ccccccl} 
    \hline
     Variant & $\gamma$ (mag) & logP<1.72 cut & filter trans.\tnote{a} & $A^{555}/A^{350}$ & trans. uncer.\tnote{b} & comments \\ \hline
       1 & $-0.035\pm0.037$  & no & R20  & R20 & no & reproduction of R20 \\
       2 & $-0.016\pm0.041$  & yes & R20 & R20 & no & \\
       3 & $0.012\pm0.041$  & yes & this work & this work, empirical &  no & \\
       4 & $0.013\pm0.057$  & yes & this work & this work, empirical &  yes & final result\\
       5 & $0.054\pm0.041$  & yes & this work & this work, speculative & no & \\
       6 & $0.055\pm0.056$  & yes & this work & this work, speculative & yes & \\
       
    \end{tabular}
    \begin{tablenotes}
        \item [a] The source of the $A^{160}/A^{H},\:A^{555}/A^{V},$ and $A^{H,\rm{MW}}/A^{V,\rm{MW}}$ transformations.
        \item [b] Inclusion of the filter transformation uncertainty. See text for details.
    \end{tablenotes}
\end{threeparttable}
\end{table*}

In order to calculate the contribution of the transformation uncertainties to the total error (not included in the R20 analysis), we inspect the change of $\gamma$ when the transformations are allowed to change within their uncertainty values. We added in quadratures the contributions from the uncertainty in $A^{H,\rm{MW}}/A^{V,\rm{MW}}$ (by scanning the uncertainty ellipse derived from the fit in Section~\ref{sec:amp_rat_HV}; $\delta\gamma\approx0.035\,\rm{mag}$), the uncertainty in $A^{555}/A^{V}$ (by changing $f$ between 0 and 1, see Appendix~\ref{sec:Ground-HST transformation}; $\delta\gamma\approx0.012\,\rm{mag}$), the uncertainty in $A^{160}/A^{H}$ (by changing $f$ between 0 and 1, see Appendix~\ref{sec:Ground-HST transformation}; $\delta\gamma\approx0.008\,\rm{mag}$), and the uncertainty in the empirical $A^{555}/A^{350}$ ($\delta\gamma\approx0.01\,\rm{mag}$). The total transformation uncertainties ($\delta\gamma\approx0.040\,\rm{mag}$) were convoluted with $\exp(-\Delta\chi^2/2)$ found without these uncertainties to increase the error in $\gamma$ from $\myapprox0.041\,\rm{mag}$ to the values presented in Table~\ref{tab:gamma_values} and in Figure~\ref{fig:gamma} (blue and green lines)\footnote{Note that the best-fit value slightly shifts because $\Delta\chi^2$ is not symmetric in $\gamma$ around the best-fit value.}. 

We claimed that comparing the extragalactic Cepheids to the MW Cepheids should be limited to $\logp<1.72$. One could worry that we ignore too many extragalactic Cepheids with this period cut, and that the period cut is too abrupt. We repeat our analysis without any period cut, but in order to reflect the more uncertain MW relation at long periods, we count for each extragalactic Cepheid the number of MW Cepheids, $N_i$, within a $0.1$ $\logp$ bin around its period, and add $\sigma_{\rm{MW}}/\sqrt{N_i}$ to its error budget, where $\sigma_{\rm{MW}}\approx0.04\,\rm{mag}$ is the intrinsic scatter of $A^{H,\rm{MW}}/A^{V,\rm{MW}}$ (see Section~\ref{sec:amp_rat_HV}). We find in this case (hereafter period weighting) an increase in $\gamma$ by only $\myapprox0.01\,\rm{mag}$. We can also use a quadratic relation for $A^{H,\rm{MW}}/A^{V,\rm{MW}}$ instead of the default linear relation (the quadratic relation is preferred over the linear relation by less than $2\sigma$, see Appendix~\ref{sec:amp_rat} for details). We find in this case a small decrease in $\gamma$ by $\myapprox0.01\,\rm{mag}$ for the $\logp<1.72$ limit case and an additional small decrease by $\myapprox0.005\,\rm{mag}$ with period weighting.  

A small fraction of the extragalactic Cepheids is found in regions with higher surface brightness (up to $\Delta m_{H}\approx2\,\rm{mag}$) than the limit typically used to measure $H_0$. We repeat our analysis by limiting the extragalactic Cepheids to small surface brightness ($\Delta m_{H}<0.7\,\rm{mag}$). We find a small increases $\delta\gamma\lesssim0.01\,\rm{mag}$. 

While the obtained $\gamma$ is consistent with the value derived by R20, the error is somewhat larger, and the best-fitting value is shifted by $\myapprox0.04\,\rm{mag}$ (for the empirical $A^{555}/A^{350}$). 
 
\section{Discussion}
\label{sec:discussion}

In this paper, we repeated the analysis of R20 to constrain a systematic blending bias, $\gamma$, through Cepheid amplitudes. The analysis compares MW Cepheids to extragalactic Cepheids, so it requires an accurate determination of Cepheid amplitudes in the MW and various filter transformations. The main differences between our analysis and the analysis of R20 are: 
\begin{enumerate}
\item We limit the extragalactic and MW Cepheids comparison to periods $\logp<1.72$, since the number of MW Cepheids with longer periods is minimal, see Appendix~\ref{sec:properties}; 
\item We use publicly available data to recalibrate amplitude ratios of MW Cepheids in standard passbands;
\item We remeasure the amplitudes of Cepheids in NGC 5584 and NGC 4258 in two \textit{HST} filters ($F555W$ and $F350LP$) to improve the empirical constraint on their amplitude ratio $A^{555}/A^{350}$.
\end{enumerate}

Our final estimates for a possible blending bias is $\gamma=0.013\pm0.057\,\rm{mag}$ with the empirical $A^{555}/A^{350}$. While the obtained $\gamma$ is consistent with the value derived by R20 and with $\gamma=0$ hence no evidence of a bias, the error is somewhat larger, and the best-fitting value is shifted by $\myapprox0.04\,\rm{mag}$. 

We constructed a galactic Cepheid catalogue from publicly available photometry for the recalibration of the MW Cepheids amplitudes ratios  (Appendix~\ref{sec:catalog}). We employed GP interpolations on the phase-folded light curves to determine the mean magnitudes and amplitudes in different bands. The GP interpolations do not depend on any presumed behaviour and allowed us to assign reliable error bars to our results. The catalogue, as well as the light curves of all Cepheids in the catalogue, are publicly available\footnote{\url{https://drive.google.com/drive/folders/1pCWp0_QARVE6EzsDSI5bMOaKdvIRHV6D?usp=sharing}}. 

We next inspect the effect of our results on the significance of the Hubble tension, by calculating $\partial H_0/\partial\gamma$ with the fitting procedure of \citet{Mortsell2021} (which is similar to the procedure of R16; a detailed description of the fitting process can be found in these papers) and the early data set release of R22. We note that some improvements to the fitting procedure and additional 18 hosts were introduced in R22, which are not included in our analysis. However, the impact of these additions should have a minor effect on $\partial H_0/\partial\gamma$. Note further that the blending bias $\gamma$ deduced from Cepheid amplitudes is actually the difference between the NIR blending bias and the white filter blending bias, $\gamma_{160}-\gamma_{350}$, such that it is not straight forward to deduce the blending bias in the Wesenheit index, $\rm{F160}-0.386(\rm{F555}-\rm{F814})$, used for the $H_0$ calculation. In what follows, we assume that the blending bias of the term $0.386(\rm{F555}-\rm{F814})$ is small compared with $\gamma_{160}$ and we take $\gamma_{350}=0$ to test the minimal effect of $\gamma$ on $H_0$ ($\gamma_{160}>\gamma$ for any positive value of $\gamma_{350}$). 

We first assume that all extragalactic Cepheids (beyond M31) are fainter by some value $\gamma$. The change in $H_0$ for the usual choice of anchors (MW, LMC and NGC 4258) is $\delta H_0/H_0\approx-0.32\gamma$ (with the same change in $H_0$ error). Limiting the bias for Cepheids with $\logp>1$, as the amplitudes observations are only available for such Cepheids, has a small effect on the results. In what follows, we assume the latest determination of the Hubble constant by the SH0ES collaboration, $H_0=73.04\pm1.04\,\rm{km}\,\rm{s}^{-1}\,\rm{Mpc}^{-1}$ (R22) and the derivative  $\delta H_0/H_0\approx-0.32\gamma$. One can now calculate the distance from the SH0ES result for any value of $\gamma$ (dashed brown lines in Figure~\ref{fig:gamma}). In order to remove the Hubble tension, a value of $\bar{\gamma}=0.24 \,\rm{mag}$ is required. At face value this gamma would seem to imply $H_0=72.7$, however it should not be interpreted that way because this method was not used to measure $H_0$; rather we conclude from it that there is no evidence of the reduced light curve amplitudes that would accompany unrecognized crowding.

A larger $\gamma$ is required to remove the tension with other combinations of anchors. For example, we find $\delta H_0/H_0\approx-0.23\gamma$ with just using the LMC and NGC 4258 anchors. The determination of $H_0$ with only the NGC 4258 anchor is hardly affected by $\gamma$ in this case, as almost all Cepheids (except M31 Cepheids) suffer from the same blending bias. We study in detail various ways to determine $H_0$ that are immune to blending biases in a companion paper \citep{Kushnir2023}.

R22 provided a few checks (see the comparison between fits 41 and 42 and the checks in Appendix B of R22) demonstrating that a possible blending is not likely related to the size of the crowding correction. A different scenario, which is more difficult to test with the methods of R22, is of a possible blending due to stars physically associated with Cepheids. Since the mass (and the age) of Cepheids is correlated with their period, it is expected that long-period Cepheids are more likely to be physically associated with stars. For example, \citet{Anderson2018} demonstrated by observing Cepheids in M31 that long period Cepheids have a higher chance of being in open clusters (see their Figure 13). The available data, however, are limited to Cepheid ages older than $\mysim50\,\rm{Myr}$ \citep[see also][with similar age limitations]{Breuval2023}. The ages of the long-period Cepheids, which dominate the population in the faraway galaxies, are $\lesssim20\,\rm{Myr}$ \citep[see, e.g, Table A1 of][]{Anderson2016}, probably shorter than the dispersing time of open clusters. It is, therefore, reasonable to assume that a significant fraction of long-period Cepheids reside in open clusters. Such an effect would lead to an increased blending with the period. We, therefore, test for such a period-dependency by modifying $\gamma$ in Equation~\eqref{eq:gamma} to $\gamma_0+\gamma_p(\logp-1)$, and repeating the fits. We find $\gamma_0=-0.28\pm0.12\,\rm{mag}$, $\gamma_p=0.61\pm0.25\,\rm{mag}\,\rm{dex}^{-1}$, with a change of $\Delta\chi^2\approx5.2$ for the empirical $A^{555}/A^{350}$, indicating an insignificant (less than $3\sigma$) evidence for a linear period dependency of $\gamma$. 

While many assumptions are involved in our analysis, we demonstrated that the R20 calibration of $\gamma=-0.029\pm0.037\,\rm{mag}$ is not secured. As we mentioned above, our results are sensitive to the $A^{555}/A^{350}$ ratio, and the empirical ratio that we use is not free from caveats. We next consider the impact of the speculative $A^{555}/A^{350}=1.15$, which represents the high-end range of (less robust) estimates. In this case we find $\gamma=0.055\pm0.056\,\rm{mag}$, which yields $H_0=71.7\,\rm{km}\,\rm{s}^{-1}\,\rm{Mpc}^{-1}$ that is $\myapprox2.5\sigma$ away from \textit{Planck}. We suggest below a few directions for future studies in order to remove some of the assumptions made in this work and to better constrain the blending effect.

We assumed that all Cepheids in NGC 4258 and the faraway galaxies suffer on average from the same systematic blending bias, which we calibrated from a smaller sample of Cepheids (and only in three faraway galaxies). Similar information for more extragalactic Cepheids can be collected with future \textit{HST} observations. Better calibration of the $A^{555}/A^{V}$, $A^{160}/A^{H}$ and $A^{555}/A^{350}$ transformations can be obtained by observations of Galactic Cepheids in many epochs, either with \textit{HST} or from the ground. Such observations could also be useful to improve existing Cepheid templates (such as P12 or the templates used by J21). For example, using the same approach of P12, but with the additional (some of them already available) \textit{HST} single epoch observations, may significantly improve the accuracy of P12 templates (that is currently estimated to be $\gtrsim10\%$). 

A different approach is to anchor the extragalactic Cepheid amplitudes to M31 Cepheids instead to the MW Cepheids. This has the advantage of observing the Cepheids with the same instrument and filters, bypassing the need for filter transformations and perhaps obtaining a larger number of long period Cepheids. Finally, the possible underline open cluster population of extragalactic Cepheids can be either examined with \textit{HST} UV observations \citep{Anderson2021} or resolved with JWST \citep{Anderson2018,Riess2021b,Yuan2022}. 

\section*{Acknowledgements}
We thank Ond\v{r}ej Pejcha, Dan Scolnic, Stefano Casertano, Eli Waxman, Boaz Katz, and Eran Ofek for useful discussions. DK is supported by the Israel Atomic Energy Commission -- The Council for Higher Education -- Pazi Foundation, by a research grant from The Abramson Family Center for Young Scientists, by ISF grant, and by the Minerva Stiftung. This research has made use of the International Variable Star Index (VSX) database, operated at AAVSO, Cambridge, Massachusetts, USA.

\section*{Data availability}

All data used in this study is either publicly available through other publications or through the publicly available catalogues:  \url{https://drive.google.com/drive/folders/1pCWp0_QARVE6EzsDSI5bMOaKdvIRHV6D?usp=sharing}. 


\begin{thebibliography}{99}
\bibitem[\protect\citeauthoryear{Anderson et al.}{2016}]{Anderson2016} Anderson R.~I., Saio H., Ekstr{\"o}m S., Georgy C., Meynet G., 2016, A\&A, 591, A8. doi:10.1051/0004-6361/201528031
\bibitem[\protect\citeauthoryear{Anderson \& Riess}{2018}]{Anderson2018} Anderson R.~I., Riess A.~G., 2018, ApJ, 861, 36. doi:10.3847/1538-4357/aac5e2
\bibitem[\protect\citeauthoryear{Anderson et al.}{2021}]{Anderson2021} Anderson R.~I., Casertano S., Riess A., Spetsieri Z., 2021, hst..prop, 16688
\bibitem[\protect\citeauthoryear{Barnes et al.}{1997}]{Barnes1997} Barnes T.~G., Fernley J.~A., Frueh M.~L., Navas J.~G., Moffett T.~J., Skillen I., 1997, PASP, 109, 645. doi:10.1086/133927
\bibitem[\protect\citeauthoryear{Berdnikov}{2008}]{Berdnikov2008} Berdnikov L.~N., 2008, yCat, II/285
\bibitem[\protect\citeauthoryear{Berdnikov et al.}{2015}]{Berdnikov2015} Berdnikov L.~N., Kniazev A.~Y., Sefako R., Dambis A.~K., Kravtsov V.~V., Zhuiko S.~V., 2015, AstL, 41, 23. doi:10.1134/S1063773715020012
\bibitem[\protect\citeauthoryear{Berdnikov et al.}{2019}]{Berdnikov2019} Berdnikov {\^A}. L. {\^A}. N., Kniazev {\^A}. A., Dambis {\^A}. A., Kravtsov {\^A}. V. {\^A}. V., 2019, PZ, 39, 2
\bibitem[\protect\citeauthoryear{Berdnikov \& Pastukhova}{2020}]{Berdnikov2020} Berdnikov L.~N., Pastukhova E.~N., 2020, AstL, 46, 235. doi:10.1134/S1063773720040027
\bibitem[\protect\citeauthoryear{Breuval et al.}{2023}]{Breuval2023} Breuval L., Riess A.~G., Macri L.~M., Li S., Yuan W., Casertano S., Konchady T., et al., 2023, arXiv, arXiv:2304.00037. doi:10.48550/arXiv.2304.00037
\bibitem[\protect\citeauthoryear{Chen et al.}{2020}]{Chen2020} Chen X., Wang S., Deng L., de Grijs R., Yang M., Tian H., 2020, ApJS, 249, 18. doi:10.3847/1538-4365/ab9cae
\bibitem[\protect\citeauthoryear{Coulson \& Caldwell}{1985}]{Coulson1985} Coulson I.~M., Caldwell J.~A.~R., 1985, SAAOC, 9
\bibitem[\protect\citeauthoryear{Di Valentino et al.}{2021}]{DiValentino2021} Di Valentino E., Mena O., Pan S., Visinelli L., Yang W., Melchiorri A., Mota D.~F., et al., 2021, CQGra, 38, 153001. doi:10.1088/1361-6382/ac086d
\bibitem[\protect\citeauthoryear{Efstathiou}{2020}]{Efstathiou2020} Efstathiou G., 2020, arXiv, arXiv:2007.10716
\bibitem[\protect\citeauthoryear{Feast et al.}{2008}]{Feast2008} Feast M.~W., Laney C.~D., Kinman T.~D., van Leeuwen F., Whitelock P.~A., 2008, MNRAS, 386, 2115. doi:10.1111/j.1365-2966.2008.13181.x
\bibitem[\protect\citeauthoryear{Fernie}{1979}]{Fernie1979} Fernie J.~D., 1979, PASP, 91, 67. doi:10.1086/130443
\bibitem[\protect\citeauthoryear{Fernie et al.}{1995}]{Fernie1995} Fernie J.~D., Evans N.~R., Beattie B., Seager S., 1995, IBVS, 4148, 1
\bibitem[\protect\citeauthoryear{Follin \& Knox}{2018}]{Follin2018} Follin B., Knox L., 2018, MNRAS, 477, 4534. doi:10.1093/mnras/sty720
\bibitem[\protect\citeauthoryear{Groenewegen}{2018}]{Groenewegen2018} Groenewegen M.~A.~T., 2018, A\&A, 619, A8. doi:10.1051/0004-6361/201833478
\bibitem[\protect\citeauthoryear{Groenewegen}{2020}]{Groenewegen2020} Groenewegen M.~A.~T., 2020, A\&A, 635, A33. doi:10.1051/0004-6361/201937060
\bibitem[\protect\citeauthoryear{Henden}{1996}]{Henden1996} Henden A.~A., 1996, AJ, 111, 902. doi:10.1086/117837
\bibitem[\protect\citeauthoryear{Hertzsprung}{1926}]{Hertzsprung1926} Hertzsprung E., 1926, BAN, 3, 115
\bibitem[\protect\citeauthoryear{Hoffmann et al.}{2016}]{Hoffmann16} Hoffmann S.~L., Macri L.~M., Riess A.~G., Yuan W., Casertano S., Foley R.~J., Filippenko A.~V., et al., 2016, ApJ, 830, 10. doi:10.3847/0004-637X/830/1/10
\bibitem[\protect\citeauthoryear{Javanmardi et al.}{2021}]{Javanmardi2021} Javanmardi B., M{\'e}rand A., Kervella P., Breuval L., Gallenne A., Nardetto N., Gieren W., et al., 2021, ApJ, 911, 12. doi:10.3847/1538-4357/abe7e5
\bibitem[\protect\citeauthoryear{Jayasinghe et al.}{2020}]{Jayasinghe2020} Jayasinghe T., Stanek K.~Z., Kochanek C.~S., Shappee B.~J., Holoien T.~W.-S., Thompson T.~A., Prieto J.~L., et al., 2020, MNRAS, 491, 13. doi:10.1093/mnras/stz2711
\bibitem[\protect\citeauthoryear{Klagyivik \& Szabados}{2009}]{Klagyivik2009} Klagyivik P., Szabados L., 2009, A\&A, 504, 959. doi:10.1051/0004-6361/200811464
\bibitem[\protect\citeauthoryear{Koen et al.}{2007}]{Koen2007} Koen C., Marang F., Kilkenny D., Jacobs C., 2007, MNRAS, 380, 1433. doi:10.1111/j.1365-2966.2007.12100.x
\bibitem[\protect\citeauthoryear{Kushnir \& Sharon }{2024}]{Kushnir2023} Kushnir D., Sharon A., 2024, in preparation
\bibitem[\protect\citeauthoryear{Laney \& Stobie}{1992}]{Laney1992} Laney C.~D., Stobie R.~S., 1992, A\&AS, 93, 93
\bibitem[\protect\citeauthoryear{Leavitt \& Pickering}{1912}]{Leavitt1912} Leavitt H.~S., Pickering E.~C., 1912, HarCi, 173
\bibitem[\protect\citeauthoryear{Moffett \& Barnes}{1984}]{Moffett1984} Moffett T.~J., Barnes T.~G., 1984, ApJS, 55, 389. doi:10.1086/190960
\bibitem[\protect\citeauthoryear{Monson \& Pierce}{2011}]{Monson2011} Monson A.~J., Pierce M.~J., 2011, ApJS, 193, 12. doi:10.1088/0067-0049/193/1/12
\bibitem[\protect\citeauthoryear{Mortsell et al.}{2021}]{Mortsell2021} Mortsell E., Goobar A., Johansson J., Dhawan S., 2021, arXiv, arXiv:2105.11461
\bibitem[\protect\citeauthoryear{Ngeow}{2012}]{Ngeow2012} Ngeow C.-C., 2012, ApJ, 747, 50. doi:10.1088/0004-637X/747/1/50
\bibitem[\protect\citeauthoryear{Pejcha \& Kochanek}{2012}]{Pejcha2012} Pejcha O., Kochanek C.~S., 2012, ApJ, 748, 107. doi:10.1088/0004-637X/748/2/107
\bibitem[\protect\citeauthoryear{Pel}{1976}]{Pel1976} Pel J.~W., 1976, A\&AS, 24, 413
\bibitem[\protect\citeauthoryear{Planck Collaboration et al.}{2020}]{Planck2020} Planck Collaboration, Aghanim N., Akrami Y., Ashdown M., Aumont J., Baccigalupi C., Ballardini M., et al., 2020, A\&A, 641, A6. doi:10.1051/0004-6361/201833910
\bibitem[\protect\citeauthoryear{Reid, Pesce, \& Riess}{2019}]{Reid2019} Reid M.~J., Pesce D.~W., Riess A.~G., 2019, ApJL, 886, L27. doi:10.3847/2041-8213/ab552d
\bibitem[\protect\citeauthoryear{Riess et al.}{2016}]{Riess2016} Riess A.~G., Macri L.~M., Hoffmann S.~L., Scolnic D., Casertano S., Filippenko A.~V., Tucker B.~E., et al., 2016, ApJ, 826, 56. doi:10.3847/0004-637X/826/1/56
\bibitem[\protect\citeauthoryear{Riess et al.}{2018}]{Riess2018} Riess A.~G., Casertano S., Yuan W., Macri L., Bucciarelli B., Lattanzi M.~G., MacKenty J.~W., et al., 2018, ApJ, 861, 126. doi:10.3847/1538-4357/aac82e
\bibitem[\protect\citeauthoryear{Riess}{2019}]{Riess2019} Riess A.~G., 2019, NatRP, 2, 10. doi:10.1038/s42254-019-0137-0
\bibitem[\protect\citeauthoryear{Riess et al.}{2020}]{Riess2020} Riess A.~G., Yuan W., Casertano S., Macri L.~M., Scolnic D., 2020, ApJL, 896, L43. doi:10.3847/2041-8213/ab9900
\bibitem[\protect\citeauthoryear{Riess et al.}{2021a}]{Riess2021a} Riess A.~G., Casertano S., Yuan W., Bowers J.~B., Macri L., Zinn J.~C., Scolnic D., 2021a, ApJL, 908, L6. doi:10.3847/2041-8213/abdbaf
\bibitem[\protect\citeauthoryear{Riess et al.}{2021b}]{Riess2021b} Riess A., Anderson R.~I., Breuval L., Casertano S., Macri L.~M., Scolnic D., Yuan W., 2021b, jwst.prop, 1685
\bibitem[\protect\citeauthoryear{Riess et al.}{2022}]{Riess2022} Riess A.~G., Yuan W., Macri L.~M., Scolnic D., Brout D., Casertano S., Jones D.~O., et al., 2022, ApJL, 934, L7. doi:10.3847/2041-8213/ac5c5b
\bibitem[\protect\citeauthoryear{Ripepi et al.}{2019}]{Ripepi2019} Ripepi V., Molinaro R., Musella I., Marconi M., Leccia S., Eyer L., 2019, A\&A, 625, A14. doi:10.1051/0004-6361/201834506
\bibitem[\protect\citeauthoryear{Rodrigo \& Solano}{2020}]{Rodrigo2020} Rodrigo C., Solano E., 2020, sea..conf, 182
\bibitem[\protect\citeauthoryear{Samus' et al.}{2017}]{Sumus2017} Samus' N.~N., Kazarovets E.~V., Durlevich O.~V., Kireeva N.~N., Pastukhova E.~N., 2017, ARep, 61, 80. doi:10.1134/S1063772917010085
\bibitem[\protect\citeauthoryear{Schechter et al.}{1992}]{Schechter1992} Schechter P.~L., Avruch I.~M., Caldwell J.~A.~R., Keane M.~J., 1992, AJ, 104, 1930. doi:10.1086/116368
\bibitem[\protect\citeauthoryear{Soszy{\'n}ski et al.}{2020}]{Soszynski2020} Soszy{\'n}ski I., Udalski A., Szyma{\'n}ski M.~K., Pietrukowicz P., Skowron J., Skowron D.~M., Poleski R., et al., 2020, AcA, 70, 101. doi:10.32023/0001-5237/70.2.2
\bibitem[\protect\citeauthoryear{Szabados}{1977}]{Szabados1977} Szabados L., 1977, CoKon, 70, 1
\bibitem[\protect\citeauthoryear{Szabados}{1980}]{Szabados1980} Szabados L., 1980, CoKon, 76, 1
\bibitem[\protect\citeauthoryear{Pietrukowicz, Soszy{\'n}ski, \& Udalski}{2021}]{Soszynski2021} Pietrukowicz P., Soszy{\'n}ski I., Udalski A., 2021, AcA, 71, 205. doi:10.32023/0001-5237/71.3.2
\bibitem[\protect\citeauthoryear{Tammann, Sandage, \& Reindl}{2003}]{Tammann2003} Tammann G.~A., Sandage A., Reindl B., 2003, A\&A, 404, 423. doi:10.1051/0004-6361:20030354
\bibitem[\protect\citeauthoryear{Turner}{2016}]{Turner2016} Turner D.~G., 2016, RMxAA, 52, 223
\bibitem[\protect\citeauthoryear{Udalski, Szyma{\'n}ski, \& Szyma{\'n}ski}{2015}]{Udalski2015} Udalski A., Szyma{\'n}ski M.~K., Szyma{\'n}ski G., 2015, AcA, 65, 1
\bibitem[\protect\citeauthoryear{Welch et al.}{1984}]{Welch1984} Welch D.~L., Wieland F., McAlary C.~W., McGonegal R., Madore B.~F., McLaren R.~A., Neugebauer G., 1984, ApJS, 54, 547. doi:10.1086/190943
\bibitem[\protect\citeauthoryear{Yoachim et al.}{2009}]{Yoachim2009} Yoachim P., McCommas L.~P., Dalcanton J.~J., Williams B.~F., 2009, AJ, 137, 4697. doi:10.1088/0004-6256/137/6/4697
\bibitem[\protect\citeauthoryear{Yuan et al.}{2022}]{Yuan2022} Yuan W., Riess A.~G., Casertano S., Macri L.~M., 2022, arXiv, arXiv:2209.09101
\end{thebibliography}

\begin{appendix}

\section{The construction of the MW catalogue}
\label{sec:catalog}

In this appendix, we describe the construction of the MW catalogue, which is used to recalibrate MW Cepheids amplitude ratios. In Section~\ref{sec:selection}, we describe the selection process of the Cepheids. In Section~\ref{sec:fits}, we present our method to determine mean magnitudes and amplitudes from publicly available photometry. In Section~\ref{sec:properties}, we discuss the content of our catalogue and determine the maximal period for which reliable results can be obtained. 

\subsection{The Cepheid selection process}
\label{sec:selection}

We aim to construct a comprehensive list of secure classical galactic Cepheids pulsating at the fundamental mode. We begin from the $1939$ fundamental mode Cepheids in the list of \citet{Soszynski2020} \citep[an updated version of the catalogue has been recently published;][which is discussed in Section~\ref{sec:properties}]{Soszynski2021}. We remove $218$ Cepheids ($64$ Cepheids with $\logp>1$) that do not have DCEP designation in the international variable star index (VSX)\footnote{\url{https://www.aavso.org/vsx/index.php}}. We add ET-Vul \citep{Berdnikov2020} and V0539-Nor to the list, with periods and positions from VSX. We finally remove from the list Cepheids that are not found in GCVS \citep{Sumus2017} or Cepheids that are identified as non-fundamental mode Cepheids by \citet{Ripepi2019}. Following this selection process, we are left with a list of $1723$ Cepheids ($424$ with $\logp>1$). 

We search the literature for high-quality, publicly available photometry of the Cepheids in our list, emphasizing Cepheids with $\logp>1$. Since the SH0ES Cepheids are observed with the $F555W$, $F814W$, and $F160W$ filters, we look for available photometry in the $V$, $I$ and $H$ bands, which are most similar to the \textit{HST} filters, respectively. Since optical (NIR) photometry sources usually include observations in the $B$ band ($J$ and $K$ bands), we include in our catalogue values for the $BVIJHK$ bands. We use the following sources for the optical photometry: \cite{Pel1976}, \cite{Szabados1977}, \cite{Szabados1980}, \cite{Moffett1984}, \cite{Coulson1985}, \cite{Henden1996}, \citet[][additional photometry is obtained from the Sternberg Astronomical Institute database\footnote{\url{http://www.sai.msu.su/groups/cluster/CEP/PHE/}}, referred later on as Bextr]{Berdnikov2008}, \cite{Berdnikov2015}, \cite{Berdnikov2019}, the OGLE Atlas of Variable Star Light Curves \citep{Udalski2015}, and the ASAS-SN Variable Stars Database \citep{Jayasinghe2020}. In the cases that the $ I $ band measurements are given in the Johnson system, we transform to the Cousins system with the transformations given in \cite{Coulson1985}. We use the following sources for the NIR photometry: \cite{Welch1984},  \cite{Laney1992}, \cite{Schechter1992}, \cite{Barnes1997}, \cite{Feast2008}, and \cite{Monson2011}. We transform the NIR photometry to the Two-Micron All-Sky Survey photometry system, using the transformations in \cite{Koen2007} and in \cite{Monson2011}. In some cases, we use the McMaster cepheid photometry and radial velocity data archive \citep{Fernie1995} to retrieve the photometry from the sources listed above.

\subsection{Light curve parameters by Gaussian processes}
\label{sec:fits}

We determine mean magnitudes and amplitudes from the retrieved photometry with interpolation using Gaussian processes (GP). The advantage of this method over template fitting methods is that it does not draw on any presumed behaviour and, for example, is not limited by intrinsic variations between light curves. The method requires sufficient sampling of the light curve, and we, therefore, require at least three epochs with the maximal phase difference between two adjacent points $<0.5$. We used the built-in \textsc{matlab} functions \textsc{fitrgp} and \textsc{predict} with a squared-exponential kernel for the covariance matrix. The phase-folded light curve is duplicated to ensure continuity, and the interpolation is performed over phases between $ -0.5 $ to $ 1.5 $. The outcome of this process is an estimated mean magnitude and an amplitude. The errors of the obtained values are estimated by repeating the process many times with the magnitude values in each phase randomly shifted according to the estimated photometric error. We choose for the photometric error the maximum between the provided observational errors (we apply a uniform error of $ 0.01\,\rm{mag} $ if no errors are provided) and the noise standard deviation, as estimated by the GP fit, which roughly corresponds to the scatter around the fit. In most cases the GP-estimated photometric error is larger than the observational photometric error, since the phase-folded light curve can have additional errors due to uncertainties in the Cepheid period (or its drift over the course of observations) or some other unknown source.

For Cepheids with $\logp>1$, we perform a consistency check of our results with the P12 templates. The templates contain the radius and temperature phase curves within the range $1\le\logp\le2$, parametrized by a truncated Fourier series, thus allowing the construction of light curves in any photometric band. We fit three parameters for each Cepheid in a given band, by minimising the deviation of the observed magnitudes, $ m_{i}^\text{obs} $, from the template-computed magnitudes at a given period, $ m_{\logp}^\text{tmp} $:
\begin{equation}\label{eq:template_fit}
\sum_i\left(m^\text{obs}_i(\phi_i)-m_{\logp}^\text{tmp}(\phi_i-\phi_0,A^2,\bar{m})\right)^2\sigma_i^{-2},
\end{equation}	
where $ \bar{m} $ is a constant offset magnitude, $ A^2 $ is the amplitude, and $ \phi_0 $ is a constant phase offset. Note that $ A^2 $ and  $ \phi_0 $ are fitted to each band separately, which is different from the method of P12, where a single value of $ A^2 $ and a single value of $ \phi_0 $ are used for all bands. We find that the differences in $ \phi_0 $ between different bands are negligible, but $ A^2 $ can change significantly between optical and NIR bands, such that a single value of $ A^2 $ for all bands is inconsistent with observations (see, for example, the deviations in $A^{H}/A^{V}$ of the P12 templates with a single value of $ A^2 $ in Section~\ref{sec:amp_rat_HV}). The advantage of template fitting over GP is the reasonable fits that are obtained for light-curves with poor sampling. However, the accuracy of the fits is limited by intrinsic variations of the light curves (at the same period) and by small scale features that are not captures by the templates. For example, the ''Hertzsprung Progression'' \citep{Hertzsprung1926}, seen as a ''bump'' in the light curves of Cepheids with periods $ \myapprox10-20\,\rm{d} $, is not captured by the templates, which leads in some cases to an underestimate of the inferred amplitudes by up to $ \myapprox0.1\,\text{mag} $. We, therefore, prefer to use the more accurate GP-derived values, but we demand that they are within $3\times\max(\sigma,0.01\,\rm{mag})$ from the templates-derived values. We further demand that the derived amplitudes are different from zero by at least $3\sigma$ (for any $\logp$).

\begin{figure*}
	\includegraphics[width=\textwidth]{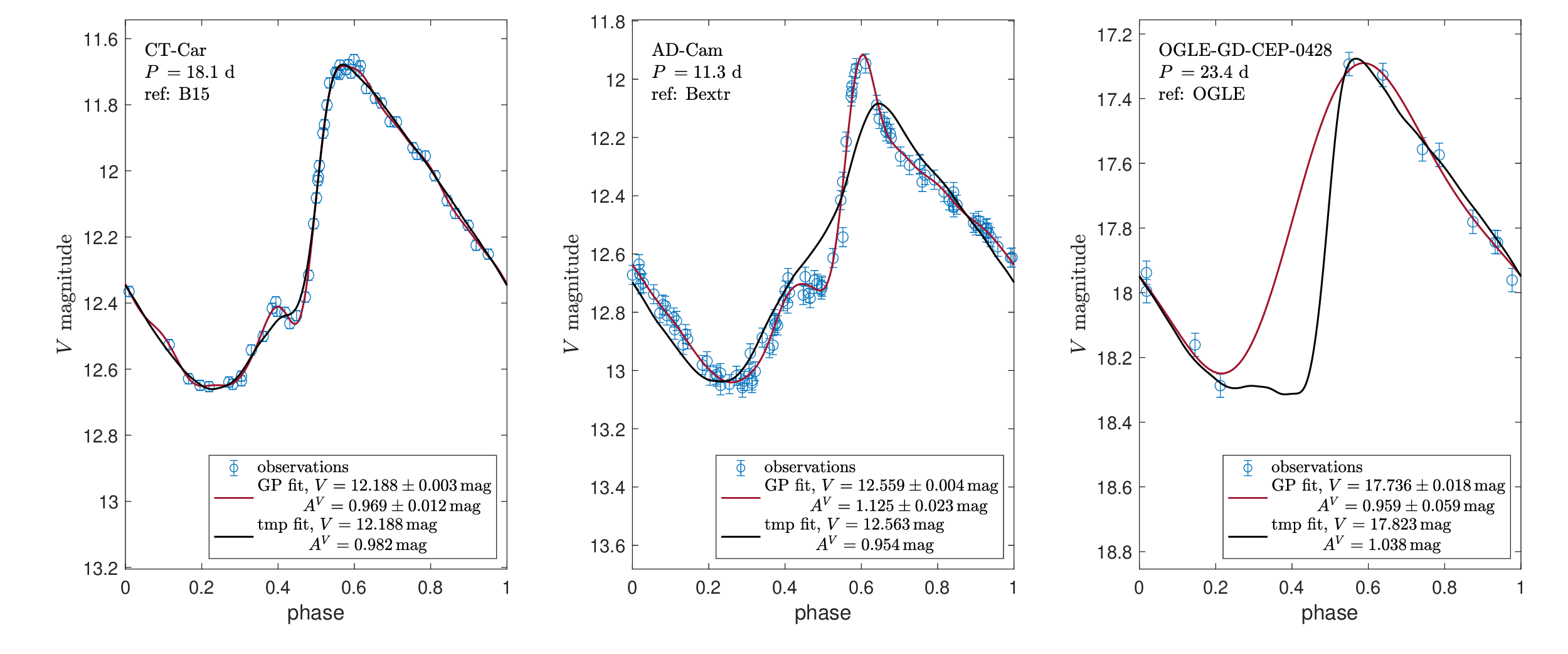}
	\includegraphics[width=0.67\textwidth]{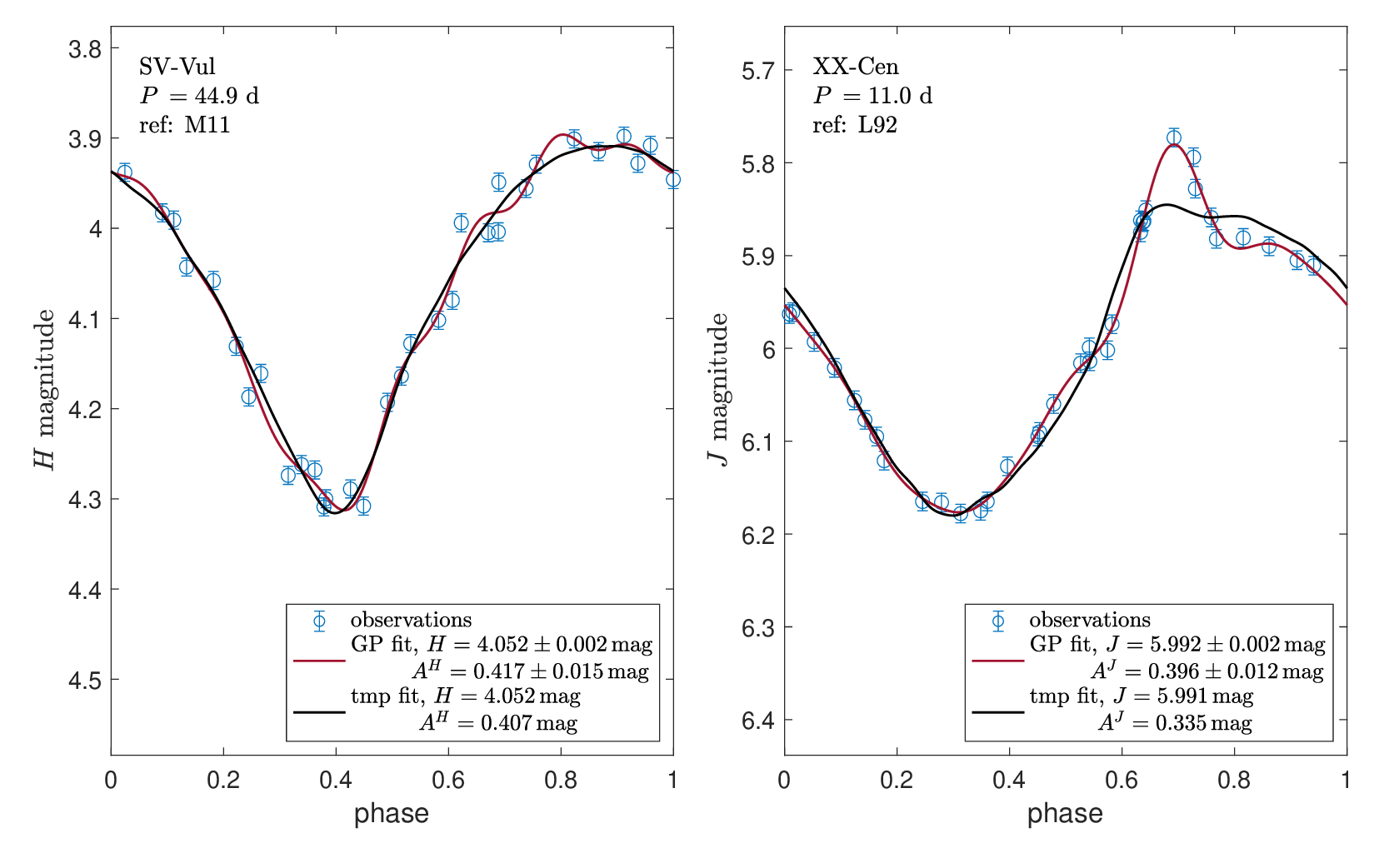}
	\caption{Examples of five Cepheid light curves and fitting results. For each Cepheid, the phase-folded observations are marked by blue circles, with error-bars corresponding to the maximum between the GP-estimated noise and the provided observational errors. The GP and template fits are indicated by the red and black lines, respectively. The three examples in the upper panel are in the $ V $ band, while the two in the lower panels are in NIR bands. CT-Car in the $V$ band and SV-Vul in the $H$ band (upper left and lower left panels, respectively) are examples of a good match between the two fits, which is the case for most of the sample. AD-Cam in the $V$ band and XX-Cen in the $J$ band (upper-middle and lower right panels, respectively) illustrate two cases where the template fits fail to capture the behaviour of the light curves, although they are well described by the GP fits. OGLE-GD-CEP-0428 in the $ V $ band (upper right panel) is a case where the GP fit significantly deviates from the template in a phase region ($0.2-0.5$) where no observations are available, and therefore the GP-derived values are rejected.}
	\label{fig:ceph_examples}
\end{figure*}

As a final check, we visually inspect all fitted light curves. Usually, the agreement between the GP-derived and the template-derived light curves is well described by our conditions from above. In Figure~\ref{fig:ceph_examples} we provide two examples for a good match between the two fits for light curves that are well sampled (CT-Car in the $V$ band, upper left panel, and SV-Vul in the $H$ band, lower left panel). In some cases, our conditions from above rejected the GP fit because the template provides a poor fit to the data. Two such examples are provided in Figure~\ref{fig:ceph_examples} (AD-Cam in the $ V $ band, upper-middle panel, and XX-Cen in the $ H $ band, lower right panel). The template fits fail to capture the behaviour of the light curves, although they are well described by the GP fits. In these cases, we keep the GP-derived values. In very few cases, our conditions from above did not reject the GP fit, although it is significantly different from the template in a phase region where no observations are available. An example is provided in Figure~\ref{fig:ceph_examples} (OGLE-GD-CEP-0428 in the $ V $ band, upper right panel). There are no observations between phases $ 0.2 $ and $ 0.5 $, where the GP fit significantly deviates from the template. In these cases, we reject the GP-derived values. Following these procedures, we obtain a catalogue that contains reliable mean magnitudes and amplitudes (with error bars) for secure classical Cepheids, especially for $\logp>1$. Figures of the obtained light curves (similar to Figure~\ref{fig:ceph_examples}) for all Cepheids in our catalogue are publicly available. 

\subsection{Properties of the catalogue}
\label{sec:properties}

Our final catalogue includes $688$ Cepheids with at least one newly derived mean magnitude or amplitude in some band. The number of newly determined mean magnitudes and amplitudes from each source is given in Tables~\ref{tab:BVI ceph} and~\ref{tab:JHK ceph} for the optical and the NIR bands, respectively. 

\begin{table*}
\begin{minipage}{120mm}
	\begin{threeparttable}
		\caption{The number of newly determined values in the catalogue from each source. Note that in some cases we reject the derived amplitude but not the derived mean magnitude. Since the data set of \citet{Berdnikov2008} and Bextr are almost identical, in many cases, the choice of source between them is arbitrary (we choose the source with smaller errors for the derived values, but the estimation of the errors contain a random component, see main text).}
		\begin{tabular}{lccccccc}
		Source & $B$ & $A^{B}$ & $V$ & $A^{V}$ & $I$ & $A^{I}$ \\ \midrule
		\cite{Pel1976} &  1 & 1 & 0 & 0 & 0 & 0 \\          
		\cite{Szabados1977} &  1 & 1 & 0 & 0 & 0 & 0 \\                     
		\cite{Szabados1980} &  1 & 1 & 1 & 1 & 0 & 0 \\           
		 \cite{Moffett1984} &  13 & 13 & 13 & 13 & 13 & 13 \\    
		 \cite{Coulson1985} &  1 & 1 & 1 & 1 & 1 & 1 \\    
		 \cite{Henden1996} &  0 & 0 & 0 & 0 & 12 & 12 \\    
		 \cite{Berdnikov2008} &  124 & 124 & 146 & 145 & 124 & 123 \\    
		 Bextr &  144 & 144 & 173 & 173 & 121 & 119 \\   
		 \cite{Berdnikov2015} &  106 & 106 & 69 & 67 & 76 & 75 \\    
		 \cite{Berdnikov2019} &  51 & 51 & 49 & 49 & 50 & 50 \\    		  
		 \cite{Udalski2015} &  0 & 0 & 36 & 28 & 223 & 220 \\    
		 \cite{Jayasinghe2020} &  0 & 0 & 10 & 10 & 0 & 0 \\    
		\end{tabular}
		\label{tab:BVI ceph}
	\end{threeparttable}
\end{minipage}
\end{table*}

\begin{table*}
\begin{minipage}{120mm}
	\begin{threeparttable}
		\caption{Same as Table~\ref{tab:BVI ceph} for the NIR bands. }
		\begin{tabular}{lcccccc}
		Source                         & $J$ & $A^{J}$ & $H$ & $A^{H}$ & $K$ & $A^{K}$\\ \midrule
		\cite{Welch1984}               &   13  &   13 &   12 & 12 &  11&  11  \\
		\cite{Laney1992}               &   32 &   32 &   31 &  31 &  31 &  31 \\
		\cite{Schechter1992}           &   14&   15 &   13 &  13 & 14 & 14\\
		\cite{Barnes1997}              &    4 &    4 &    4 &  4 &    4 &    4\\
		\cite{Feast2008}               &    5 &    5 &    5 &  5 &   5  &   5 \\		
		\cite{Monson2011}              &  126 &  126 &  126 & 126 & 123 & 122 \\
		\end{tabular}
		\label{tab:JHK ceph}
	\end{threeparttable}
	\end{minipage}
\end{table*}

\begin{figure*}
	\includegraphics[width=1\textwidth]{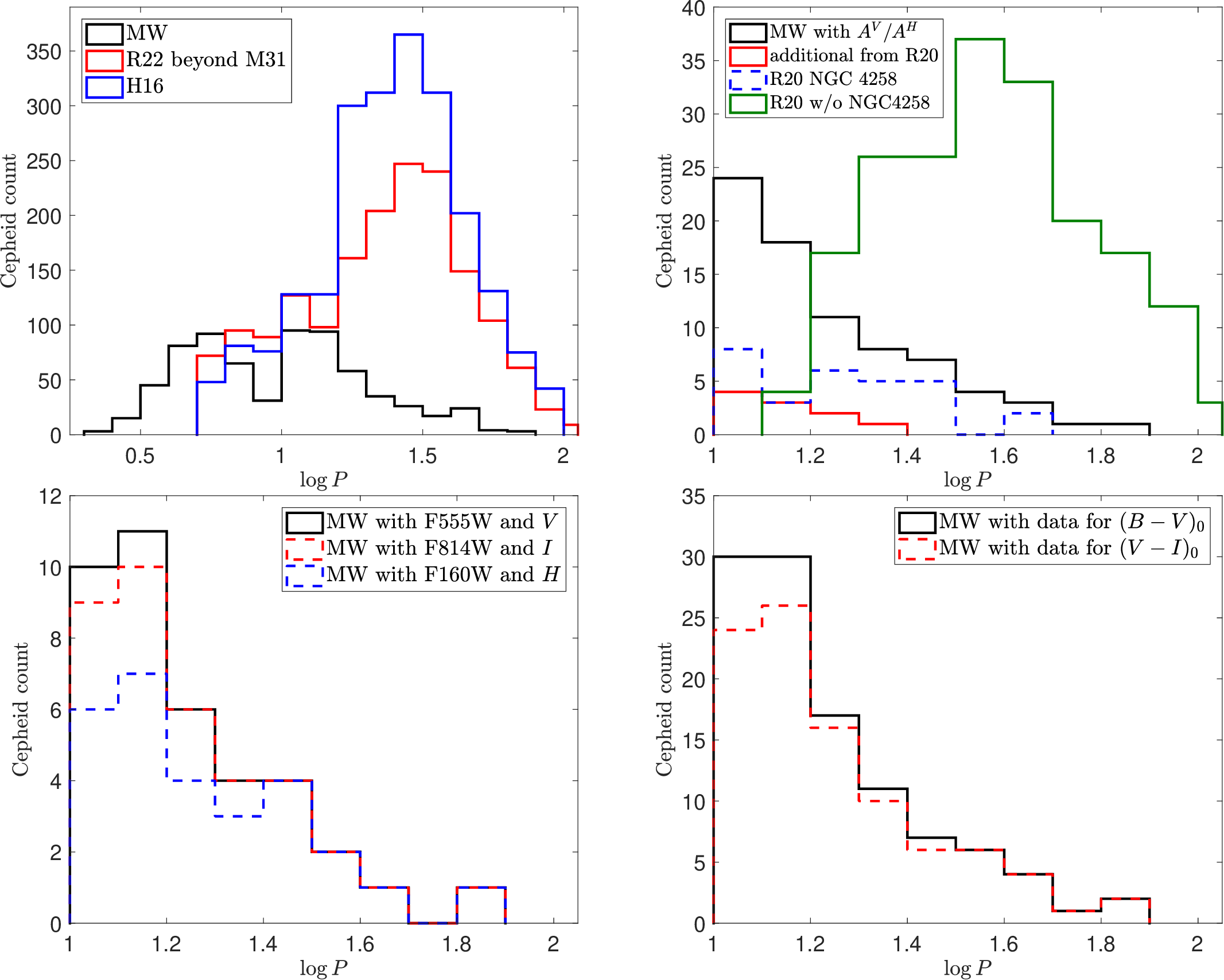}
	\caption{The distribution of periods in the catalogue with $\logp$ bin widths of $0.1$. Upper left panel: The entire catalogue (black), which contains 356 (332) Cepheids with $\logp>1$ ($\logp<1$). We demonstrate with the H16 optical sample (blue) and the R22 early data release NIR sample (red) that the vast majority of available extragalactic Cepheids for which crowding corrections are significant (i.e., beyond M31) have $\logp>1$. Upper right panel: Cepheids with both $A^{V}$ and $A^{H}$ (black). There are only two Cepheids with $\logp>1.7$, such that one cannot reliably determine the $A^{V}/A^{H}$ MW ratio in this period range. One of the two Cepheids is GY-Sge with $\logp\approx1.71$, so we set our default upper limit to be $ \logp=1.72$ to include the largest reasonable period range. Table 1 of R20 includes additional $10$ Cepheids (red) with $A^{H}$ values from unpublished photometry. To keep our data homogeneous, we do not include the reported $A^{H}$ values of these additional Cepheids. Since these Cepheids have $\logp<1.4$, where we have numerous Cepheids, the impact of ignoring these Cepheids is minimal. The R20 extragalactic sample is shown as well (in blue for NGC 4258 and in green for the faraway galaxies). All Cepheids in NGC4258 have $\logp<1.7$, but there is a significant fraction of Cepheids in the faraway galaxies with $\logp>1.7$ which cannot be reliably compared to the MW. Bottom left panel: Cepheids with both \textit{HST} ($F555W$, $F814W$, and $F160W$) and ground observations ($V$, $I$, and $H$) are shown in black, red, and blue, respectively. There are only 4 Cepheids with $\logp>1.5$, which limits the reliability of the filter transformations in this period range. Bottom right panel: Cepheids with sufficient data to determine $(B-V)_0$ (black) and $(V-I)_0$ (red). There are only three such Cepheids with $\logp>1.7$, limiting the intrinsic colour's reliability in this period range.}
	\label{fig:Count}
\end{figure*}

The distribution of periods in the catalogue is shown in the upper left panel of Figure~\ref{fig:Count}. The catalogue contains 356 (332) Cepheids with $\logp>1$ ($\logp<1$). The vast majority of available extragalactic Cepheids for which crowding corrections are significant (i.e., beyond M31) have $\logp>1$, see the upper left panel of Figure~\ref{fig:Count}. As a result, in what follows, we do not consider the short-period ($\logp<1$) Cepheids, although we provide in our catalogue their derived values.  

The number of Cepheids in our catalogue with both $A^{V}$ and $A^{H}$ (and $\logp>1$) is 77. The period distribution of these Cepheids is shown in the upper right panel of Figure~\ref{fig:Count}. As can be seen in the figure, there are only two Cepheids with $\logp>1.7$, such that one cannot reliably determine the $A^{V}/A^{H}$ MW ratio in this period range. One of the two Cepheids is GY-Sge with $\logp\approx1.71$, so we set our default upper limit to be $\logp=1.72$ to include the largest reasonable period range. Table 1 of R20 includes additional $10$ Cepheids\footnote{DR-Vel, KK-Cen, SS-CMa, XY-Car, SY-Nor, SV-Vel, XX-Car, XZ-Car, YZ-Car, and V0340-Ara.} (red histogram) with $A^{H}$ values from unpublished photometry, and are therefore not included in our catalogue. To keep our data homogeneous, we do not include the reported $A^{H}$ values of these additional Cepheids in what follows. Since these Cepheids have $\logp<1.4$, where we have numerous Cepheids, the impact of ignoring these Cepheids is minimal. The period distribution of the R20 extragalactic sample is shown as well (in blue for NGC 4258 and in green for the faraway galaxies). As can be seen, all Cepheids in NGC4258 have $\logp<1.7$, but there is a significant fraction of Cepheids in the faraway galaxies with $\logp>1.7$ which cannot be reliably compared to the MW. 

\vspace{5mm}

We supplement the catalogue with \textit{HST} observations in the $F555W$, $F814W$, and $F160W$ filters, as reported by \citet{Riess2018,Riess2021a}. This data can be used to determine various transformations between \textit{HST} and ground filters. The bottom left panel of Figure~\ref{fig:Count} shows the period distribution of Cepheids with both \textit{HST} and ground observations. As can be seen, there are only four Cepheids with $\logp>1.5$, which limits the reliability of the filter transformations in this period range. We finally provide selective extinction, $E(B-V)$, values that can be used to calculate the intrinsic colours of Cepheids. The preferred source for $E(B-V)$ values is \citet{Turner2016}, with additional values (in order of preference) from \citet{Groenewegen2020,Fernie1995,Ngeow2012}. We multiply the estimates of \citet{Fernie1995} by $0.94$ \citep[see discussion in][]{Tammann2003,Groenewegen2018}. The bottom right panel of Figure~\ref{fig:Count} shows the period distribution of Cepheids with sufficient data to determine various intrinsic colours. As can be seen, there are only three such Cepheids with $\logp>1.7$, which limits the reliability of the intrinsic colour in this period range.

\newgeometry{margin=1cm}
    \begin{landscape}
    
\begin{table}
\caption{A few examples for entries in the catalogue. For each Cepheid we provide position (RA and Dec in degrees), period (in d), source for the position and period (S20 \citep{Soszynski2020} or VSX), and the fitting results (mean magnitude and amplitude in each band together with the source of photometry). The photometry sources initials are: P76 \citep{Pel1976}, S77 \citep{Szabados1977}, S80 \citep{Szabados1980}, M84 \citep{Moffett1984}, C85 \citep{Coulson1985}, H96 \citep{Henden1996}, B08 \citep{Berdnikov2008}, Bextr (photometry from the Sternberg Astronomical Institute database), B15 \citep{Berdnikov2015}, B19 \citep{Berdnikov2019}, OGLE \citep{Udalski2015}, ASAS \citep{Jayasinghe2020}.}
\begin{tabular}{|c||c||c||c||c||c||c||c||c||c||c||c||c||c|}
\hline
Name & RA & Dec & Period & Source & $B$ & $A^{B}$ & ref & $V$ & $A^{V}$ & Ref & $I$ & $A^{I}$ & ref   \\ 
  \hline
FF-Aur	&	73.82550	&	39.97975	&	2.121	&	OF	&	$14.687\pm0.006$	&	$1.440\pm0.027$	&	B08	&	$13.720\pm0.004$	&	$1.075\pm0.019$	&	Bextr	&	$12.455\pm0.028$	&	$0.501\pm0.149$	&	H96	\\
     BB-Gem	&	98.64708	&	13.07911	&	2.308	&	OF	&	$12.244\pm0.015$	&	$1.440\pm0.055$	&	B15	&	$11.412\pm0.011$	&	$1.054\pm0.041$	&	B15	&	$10.416\pm0.007$	&	$0.669\pm0.029$	&	B15	\\
     V2201-Cyg	&	316.07011	&	49.74440	&	2.418	&	OF	&	$13.722\pm0.006$	&	$0.751\pm0.021$	&	Bextr	&	$12.126\pm0.002$	&	$0.532\pm0.009$	&	Bextr	&	$99.900\pm99.900$	&	$99.900\pm99.900$	&		\\
     CN-CMa	&	107.39410	&	-18.56329	&	2.546	&	OF	&	$99.900\pm99.900$	&	$99.900\pm99.900$	&		&	$13.664\pm0.002$	&	$0.750\pm0.009$	&	ASAS	&	$12.240\pm0.018$	&	$0.581\pm0.042$	&	B08	\\
     XZ-CMa	&	105.10346	&	-20.43169	&	2.558	&	OF	&	$13.761\pm0.006$	&	$1.448\pm0.022$	&	B08	&	$12.933\pm0.004$	&	$1.029\pm0.014$	&	Bextr	&	$11.904\pm0.004$	&	$0.637\pm0.014$	&	Bextr	\\
     V0620-Pup	&	119.45788	&	-29.38406	&	2.586	&	OF	&	$12.967\pm0.007$	&	$0.778\pm0.024$	&	B19	&	$11.959\pm0.004$	&	$0.541\pm0.017$	&	B19	&	$10.730\pm0.003$	&	$0.325\pm0.012$	&	B19	\\
     IT-Lac	&	332.32734	&	51.40539	&	2.632	&	OF	&	$16.046\pm0.014$	&	$0.953\pm0.054$	&	B08	&	$15.156\pm0.012$	&	$0.643\pm0.059$	&	B08	&	$99.900\pm99.900$	&	$99.900\pm99.900$	&		\\
     BW-Gem	&	93.99954	&	23.74750	&	2.635	&	OF	&	$12.991\pm0.025$	&	$1.123\pm0.084$	&	B15	&	$11.975\pm0.018$	&	$0.802\pm0.068$	&	B15	&	$10.711\pm0.013$	&	$0.499\pm0.049$	&	B15	\\
     V0539-Nor	&	245.22592	&	-53.55461	&	2.644	&	VSX	&	$12.317\pm0.004$	&	$0.513\pm0.016$	&	B19	&	$11.826\pm0.003$	&	$0.421\pm0.015$	&	B19	&	$11.208\pm0.003$	&	$0.332\pm0.010$	&	B19	\\
     EW-Aur	&	72.85342	&	38.18856	&	2.660	&	OF	&	$14.595\pm0.009$	&	$1.107\pm0.037$	&	Bextr	&	$13.515\pm0.006$	&	$0.784\pm0.029$	&	Bextr	&	$99.900\pm99.900$	&	$99.900\pm99.900$	&		\\
    . & & & & & & & & & & & & & \\
     . & & & & & & & & & & & & & \\
      V1467-Cyg	&	301.00912	&	32.45075	&	48.677	&	OF	&	$15.974\pm0.005$	&	$1.435\pm0.022$	&	Bextr	&	$13.485\pm0.003$	&	$1.006\pm0.014$	&	Bextr	&	$10.545\pm0.010$	&	$0.740\pm0.059$	&	Bextr	\\
     OGLE-GD-CEP-1499	&	288.47750	&	11.95300	&	49.142	&	OF	&	$99.900\pm99.900$	&	$99.900\pm99.900$	&		&	$99.900\pm99.900$	&	$99.900\pm99.900$	&		&	$15.154\pm0.001$	&	$0.643\pm0.006$	&	OGLE	\\
     CE-Pup	&	123.53350	&	-42.56817	&	49.322	&	OF	&	$13.379\pm0.003$	&	$1.282\pm0.012$	&	B15	&	$11.755\pm0.003$	&	$0.841\pm0.011$	&	B15	&	$9.965\pm0.003$	&	$0.532\pm0.011$	&	Bextr	\\
     OGLE-GD-CEP-1505	&	288.60517	&	12.99211	&	50.604	&	OF	&	$99.900\pm99.900$	&	$99.900\pm99.900$	&		&	$99.900\pm99.900$	&	$99.900\pm99.900$	&		&	$15.794\pm0.003$	&	$0.538\pm0.013$	&	OGLE	\\
     V0708-Car	&	153.90787	&	-59.55131	&	51.414	&	OF	&	$14.456\pm0.002$	&	$0.762\pm0.008$	&	B19	&	$12.075\pm0.002$	&	$0.536\pm0.008$	&	B19	&	$9.221\pm0.002$	&	$0.401\pm0.008$	&	B19	\\
     GY-Sge	&	293.80679	&	19.20239	&	51.814	&	OF	&	$12.445\pm0.007$	&	$1.009\pm0.019$	&	B08	&	$10.154\pm0.004$	&	$0.646\pm0.014$	&	Bextr	&	$7.529\pm0.011$	&	$0.350\pm0.024$	&	B08	\\
     ET-Vul	&	293.83375	&	26.43022	&	53.910	&	VSX	&	$13.846\pm0.005$	&	$0.689\pm0.014$	&	B08	&	$12.190\pm0.003$	&	$0.449\pm0.008$	&	Bextr	&	$99.900\pm99.900$	&	$99.900\pm99.900$	&		\\
     II-Car	&	162.20437	&	-60.06306	&	64.836	&	OF	&	$14.830\pm0.006$	&	$1.353\pm0.028$	&	B15	&	$12.580\pm0.003$	&	$0.917\pm0.015$	&	B15	&	$9.858\pm0.004$	&	$0.560\pm0.014$	&	Bextr	\\
     V1496-Aql	&	283.74804	&	-0.07678	&	65.731	&	OF	&	$99.900\pm99.900$	&	$99.900\pm99.900$	&		&	$10.185\pm0.007$	&	$99.900\pm99.900$	&	B15	&	$7.743\pm0.006$	&	$0.345\pm0.014$	&	B15	\\
     S-Vul	&	297.09921	&	27.28650	&	68.651	&	OF	&	$10.851\pm0.003$	&	$0.912\pm0.014$	&	B08	&	$8.965\pm0.002$	&	$0.552\pm0.012$	&	Bextr	&	$6.939\pm0.008$	&	$0.384\pm0.020$	&	Bextr	\\
\hline
\end{tabular}
\label{tbl:cat1}
\end{table}
	
\begin{table}
\caption{Same as Table~\ref{tbl:cat1} for NIR photometry and $E(B-V)$ values. The photometry sources initials are: W84 \citep{Welch1984}, L92 \citep{Laney1992}, S92 \citep{Schechter1992}, B97b \citep{Barnes1997}, F08 \citep{Feast2008}, M11 \citep{Monson2011}. The $E(B-V)$ sources initials are: F95 \citep{Fernie1995}, N12 \citep{Ngeow2012}, T16 \citep{Turner2016}, G20 \citep{Groenewegen2020}.}
\begin{tabular}{|c||c||c||c||c||c||c||c||c||c||c||c||c||c|}
\hline
Name & Period & source & $J$ & $A^{J}$ & Ref & $H$ & $A^{H}$ & Ref & $K$ & $A^{K}$ & Ref & $E(B-V)$ & Ref \\ 
  \hline
      . & & & & & & & & & & & & &\\
      V1467-Cyg	&	48.677	&	OF	&	$8.164\pm0.003$	&	$0.462\pm0.015$	&	M11	&	$7.286\pm0.014$	&	$0.380\pm0.050$	&	M11	&	$6.918\pm0.005$	&	$0.375\pm0.013$	&	M11	&	$1.532\pm99.900$	&	F95	\\
     OGLE-GD-CEP-1499	&	49.142	&	OF	&	$99.900\pm99.900$	&	$99.900\pm99.900$	&		&	$99.900\pm99.900$	&	$99.900\pm99.900$	&		&	$99.900\pm99.900$	&	$99.900\pm99.900$	&		&	$99.900\pm99.900$	&		\\
     CE-Pup	&	49.322	&	OF	&	$99.900\pm99.900$	&	$99.900\pm99.900$	&		&	$99.900\pm99.900$	&	$99.900\pm99.900$	&		&	$99.900\pm99.900$	&	$99.900\pm99.900$	&		&	$0.740\pm0.070$	&	G20	\\
     OGLE-GD-CEP-1505	&	50.604	&	OF	&	$99.900\pm99.900$	&	$99.900\pm99.900$	&		&	$99.900\pm99.900$	&	$99.900\pm99.900$	&		&	$99.900\pm99.900$	&	$99.900\pm99.900$	&		&	$99.900\pm99.900$	&		\\
     V0708-Car	&	51.414	&	OF	&	$99.900\pm99.900$	&	$99.900\pm99.900$	&		&	$99.900\pm99.900$	&	$99.900\pm99.900$	&		&	$99.900\pm99.900$	&	$99.900\pm99.900$	&		&	$99.900\pm99.900$	&		\\
     GY-Sge	&	51.814	&	OF	&	$5.540\pm0.003$	&	$0.273\pm0.013$	&	M11	&	$4.836\pm0.005$	&	$0.237\pm0.017$	&	M11	&	$4.520\pm0.003$	&	$0.239\pm0.010$	&	M11	&	$1.147\pm0.020$	&	T16	\\
     ET-Vul	&	53.910	&	VSX	&	$99.900\pm99.900$	&	$99.900\pm99.900$	&		&	$99.900\pm99.900$	&	$99.900\pm99.900$	&		&	$99.900\pm99.900$	&	$99.900\pm99.900$	&		&	$99.900\pm99.900$	&		\\
     II-Car	&	64.836	&	OF	&	$99.900\pm99.900$	&	$99.900\pm99.900$	&		&	$99.900\pm99.900$	&	$99.900\pm99.900$	&		&	$99.900\pm99.900$	&	$99.900\pm99.900$	&		&	$1.372\pm99.900$	&	F95	\\
     V1496-Aql	&	65.731	&	OF	&	$99.900\pm99.900$	&	$99.900\pm99.900$	&		&	$99.900\pm99.900$	&	$99.900\pm99.900$	&		&	$99.900\pm99.900$	&	$99.900\pm99.900$	&		&	$99.900\pm99.900$	&		\\
     S-Vul	&	68.651	&	OF	&	$5.412\pm0.004$	&	$0.234\pm0.012$	&	M11	&	$4.816\pm0.004$	&	$0.225\pm0.011$	&	M11	&	$4.551\pm0.006$	&	$0.224\pm0.025$	&	M11	&	$0.999\pm0.010$	&	T16	\\
    \hline
\end{tabular}
\label{tbl:cat2}
\end{table}

\end{landscape}
\restoregeometry

All Cepheids in our final catalogue are classified as fundamental mode Cepheids in the updated catalogue of \citet{Soszynski2021}. There are additional 333 Cepheids (57 with $\logp>1$) in the updated catalogue of \citet{Soszynski2021} that are not present in \citet{Soszynski2020}. The additional $\logp>1$ Cepheids are mostly from \citet{Chen2020}. We could not find NIR observations of the additional $\logp>1$ Cepheids, such that there is no available additional data that could modify the main results of this paper. 

The entire catalogue is available online. A few examples for entries in the catalogue are given in Tables~\ref{tbl:cat1}-\ref{tbl:cat2}.

\section{The amplitude ratios of the MW Cepheids}
\label{sec:amp_rat}

In this appendix, we use our catalogue to derive the amplitude ratios of the MW Cepheids with $1<\logp<1.72$ in different bands. We present in Figure~\ref{fig:AmpRatio} the ratios that are used to estimate the ground-HST filter transformations in Appendix~\ref{sec:Ground-HST transformation}. As can be seen in the top-left panel, we fit the ratio $A^{V}/A^{B}$ with a quadratic function (black line). We find in this case $\chi^2_{\nu}\approx3.7$ for $123$ Cepheids after the removal of the outliers GQ-Vul and SU-Cru, suggesting an intrinsic scatter of $\myapprox0.018$. The results of the fit following the addition of the calibrated intrinsic scatter are indicated in the figure. We find no significant improvement for fitting with a third-order polynomial (but we do find a significant improvement over fitting with a constant ratio or a linear function). The templates of P12 (red line) reproduce the fitted function with deviations $\lesssim5\%$. 

As can be seen in the bottom-left panel, we fit the ratio $A^{I}/A^{V}$ with a constant ratio (black line). We find in this case $\chi^2_{\nu}\approx4.7$ for $132$ Cepheids after the removal of the outliers OGLE-GD-CEP-0332 and SU-Cru, suggesting an intrinsic scatter of $\myapprox0.031$. The results of the fit following the addition of the calibrated intrinsic scatter are indicated in the figure. We find no significant improvement for fitting with higher-order polynomials. The templates of P12 (red line) reproduce the fitted value with deviations $\lesssim10\%$. The result of J21 for $A^{814}/A^{555}$ (dotted blue line, derived from their equation 3) is similar to our fit, however, their results should be multiplied by $(A^{I}/A^{814})(A^{555}/A^{V})$ for comparing to $A^{I}/A^{V}$. The factor $(A^{I}/A^{814})(A^{555}/A^{V})$ is estimated in Appendix~\ref{sec:Ground-HST transformation} (along with an argument for this factor exceeding $1$) and the result of multiplying this factor by the J21 result is plotted is solid blue line (the dashed blue lines represent the estimated error of this factor). The obtained $A^{I}/A^{V}$ based on the J21 result overpredict our fitted value by  $\gtrsim10\%$. We discuss in detail the J21 method in Section~\ref{sec:HST amplitude}. 

As can be seen in the top-right panel, we fit the ratio $A^{H}/A^{J}$ with a quadratic function (black line). We find in this case $\chi^2_{\nu}\approx3.7$ for $74$ Cepheids after the removal of the outlier AA-Gem, suggesting an intrinsic scatter of $\myapprox0.067$. The results of the fit following the addition of the calibrated intrinsic scatter are indicated in the figure. We find no significant improvement for fitting with a third-order polynomial (but we do find a significant improvement over fitting with a constant ratio or a linear function). The P12 templates (red line) mostly overpredict the fitted function with deviations smaller than $\myapprox15\%$. As can be seen in the bottom-right panel, we fit the ratio $A^{K}/A^{H}$ with a constant ratio (black line). We find a good fit in this case, $\chi^2_{\nu}\approx0.79$ for $73$ Cepheids after the removal of the outliers RY-Cas and YZ-Aur, suggesting that the intrinsic scatter is smaller than the observational error (the scatter of the observed ratios is $\myapprox0.07$). The results of the fit are indicated in the figure. We find no significant improvement for fitting with a linear function. The P12 templates (red line) slightly overpredict the fitted value with deviations $\lesssim5\%$. 

\begin{figure*}
	\includegraphics[width=1\textwidth]{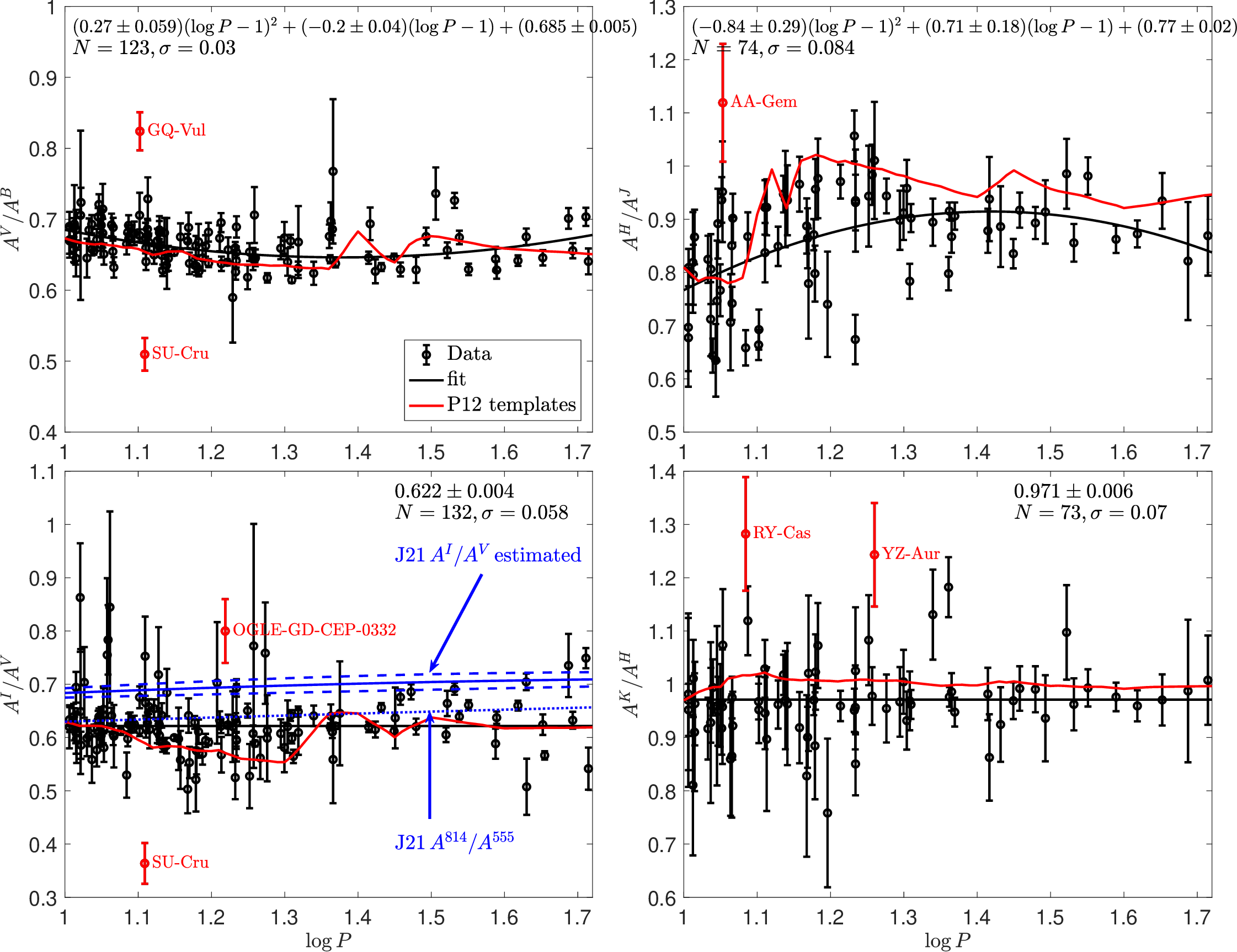}
	\caption{Distributions of amplitude ratios as a function of the period. Top-left panel: the ratio $A^{V}/A^{B}$. We fit the observations with a quadratic function (black line). The templates of P12 (red line) reproduce the fitted function with deviations $\lesssim5\%$. Bottom-left panel: the ratio $A^{I}/A^{V}$. We fit the observations with a constant ratio (black line). The templates of P12 (red line) reproduce the fitted value with deviations $\lesssim10\%$. The result of J21 for $A^{814}/A^{555}$ (dotted blue line) is similar to our fit. The obtained $A^{I}/A^{V}$ based in the J21 result (solid blue line with estimated errors in dashed blue lines) overpredict our fitted value by $\gtrsim10\%$. Top-right panel: the ratio $A^{H}/A^{J}$. We fit the observations with a quadratic function (black line). The P12 templates (red line) overpredict the fitted function with deviations smaller than $\myapprox20\%$. Bottom-right panel: the ratio $A^{K}/A^{H}$. We fit the observations with a constant ratio (black line). The P12 templates (red line) slightly overpredict the fitted value with deviations $\lesssim5\%$.}
	\label{fig:AmpRatio}
\end{figure*}

We reproduced the known result that the $\gtrsim0.1\,\rm{mag}$ scatter seen in single-band amplitudes can in some cases be significantly reduced by considering amplitude ratios between different bands \citep[][and references therein]{Klagyivik2009}. This was the motivation of R20 to study the ratio $A^{H}/A^{V}$ (see Section~\ref{sec:amp_rat_HV}).

\section{Ground-HST amplitude transformations}
\label{sec:Ground-HST transformation}

In this appendix, we estimate the ground-to-HST amplitude ratios $A^{555}/A^{V}$ and $A^{160}/A^{H}$, which are required for comparing the MW amplitudes to the extragalactic amplitudes in Section~\ref{sec:amp_extra} (see Equation~\eqref{eq:ratio transformation}), and the ratio $A^{814}/A^{I}$ (not required for our analysis). Since complete light curves of the same Cepheids with both ground and \textit{HST} filters are unavailable, we are unable to directly calibrate the required ratios (see Appendix~\ref{sec:amp_rat} for a direct calibration of other bands). We are, therefore, forced to make some approximations to estimate the required ratios. We suggest in Section~\ref{sec:discussion} future observations that will allow a more direct calibration. 

The method of R20 to estimate $A^{Z}/A^{X}$, where $Z$ is an \textit{HST} filter ($F555W$, $F814W$ or $F160W$) that is similar to a ground filter $X$ ($V$, $I$ or $H$, respectively) is as follows. They first calibrate mean-magnitude transformations in the form of 
\begin{equation}\label{eq:XYZ tran}
Z=X+zp+b(X-Y),
\end{equation}
where $Y$ ($I$, $V$ or $J$, respectively) is a nearby filter, $zp$ is the zero point, and $b$ is the slope of the colour term. They next assume that the transformation holds in each phase of the light curve and that the extremum values of the $X$, $Y$ and $Z$ light curves are at the same phase. Then they can derive the amplitude ratio $A^{Z}/A^{X}$ as
\begin{equation}\label{eq:AZ/AX}
\frac{A^{Z}}{A^{X}}=1+b-b\frac{A^{Y}}{A^{X}}.
\end{equation}
In reality, none of the assumptions from above hold, and the level at which Equation~\eqref{eq:AZ/AX} deviates from the actual ratio is difficult to estimate. Note that Equation~\eqref{eq:AZ/AX} depends only on $b$, while $b$ is highly degenerate with $zp$. In other words, there is a range of $b$ values that is consistent with the mean magnitude transformation (through degeneracy with $zp$) and significantly changes the amplitude ratio transformation. 

Here we choose to use a different method, which relies on the empirical observation that for a given Cepheid the amplitude is a decreasing function of the observed wavelength \citep{Fernie1979,Klagyivik2009}. In Figure~\ref{fig:AYoAV} we show $A^{Y}/A^{V}$ for the filters $Y=BVIJHK$, as calibrated in Appendix~\ref{sec:amp_rat}, as a function of $1/\lambda$, where $\lambda$ is the effective wavelength of filter $Y$ \citep[obtained from the SVO filter profile service;][]{Rodrigo2020}\footnote{\url{http://svo2.cab.inta-csic.es/theory/fps/}. We use $\lambda=0.443,0.554,0789,1.235,1.662,2.159\,\mu \rm{m}$ for $Y=BVIJHK$ and $\lambda=0.539,0.813,1.544\,\mu \rm{m}$ for $Z=$$F555W$,$F814W$,$F160W$.}. The motivation to use $1/\lambda$  is the linear relation that is obtained in the optical and the UV bands \citep{Fernie1979,Klagyivik2009}. As can be seen in the figure, the function $A^{Y}/A^{V}$ is super-linear with $1/\lambda$, such that estimating $A^{555}/A^{V}$ by interpolating between the $V$ and $B$ bands is expected to overestimate the ratio, while extrapolating with the $I$ and the $V$ bands is expected to underestimate the ratio. We can therefore bound $A^{555}/A^{V}$ between these two estimates. A similar bound can be obtained for $A^{160}/A^{H}$ ($A^{814}/A^{I}$) by considering $A^{Y}/A^{H}$ ($A^{Y}/A^{I}$),  interpolation with the $J$ ($J$) band, and extrapolation with the $K$ ($V$) band. Because $F555W$, $F814W$ and $F160W$ are close to the $V$, $I$ and $H$ band, respectively, our choice of using $1/\lambda$ (instead of $\lambda$, for example) has a small effect on our results.

\begin{figure}
	\includegraphics[width=\columnwidth]{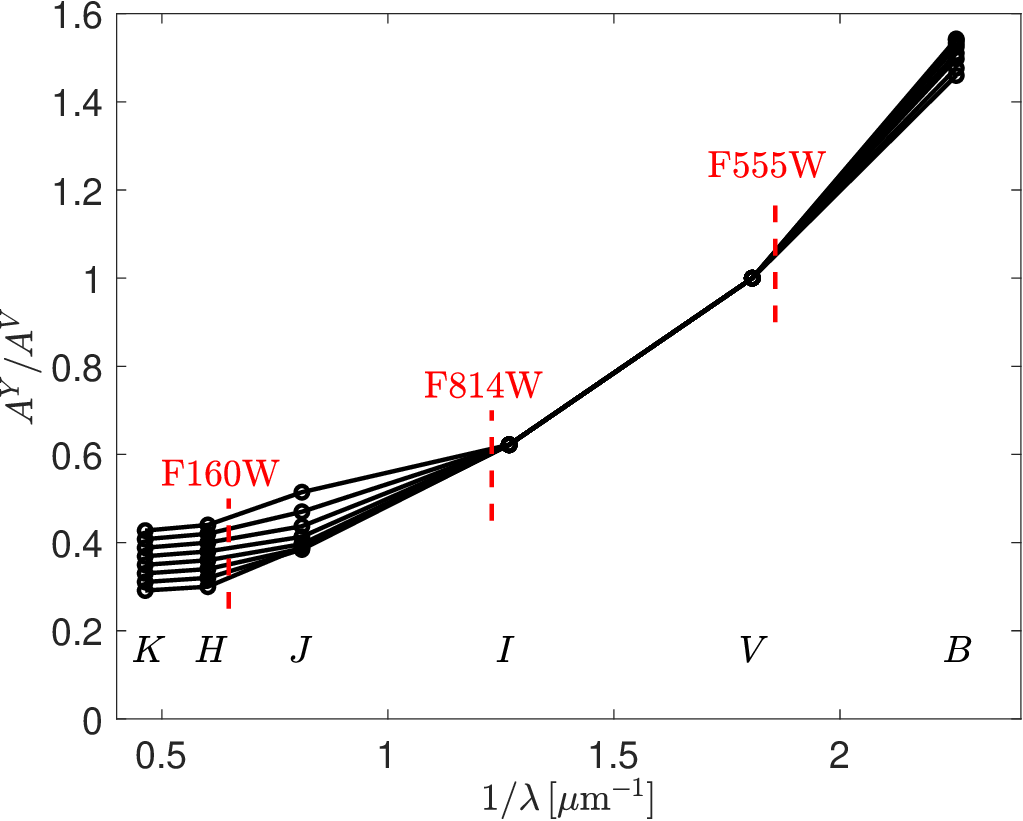}
	\caption{$A^{Y}/A^{V}$ for the filters $Y=BVIJHK$ (calibrated in Appendix~\ref{sec:amp_rat}) as a function of $1/\lambda$, where $\lambda$ is the effective wavelength of filter $Y$. Each solid line corresponds to a single value of $\logp$ within the range $[1,1.7]$ with a spacing of $0.1$. The function $A^{Y}/A^{V}$ is super-linear with $1/\lambda$, such that estimating $A^{555}/A^{V}$ by interpolating between the $V$ and $B$ bands is expected to overestimate the ratio, while extrapolating with the $I$ and the $V$ bands is expected to underestimate the ratio. We can therefore bound $A^{555}/A^{V}$ between these two estimates. A similar bound can be obtained for $A^{160}/A^{H}$ ($A^{814}/A^{I}$) by considering $A^{Y}/A^{H}$ ($A^{Y}/A^{I}$),  interpolation with the $J$ ($J$) band, and extrapolation with the $K$ ($V$) band. Our choice of using $1/\lambda$ (instead of $\lambda$, for example) has a small effect on our results.}
	\label{fig:AYoAV}
\end{figure}

\begin{figure}
	\includegraphics[width=\columnwidth]{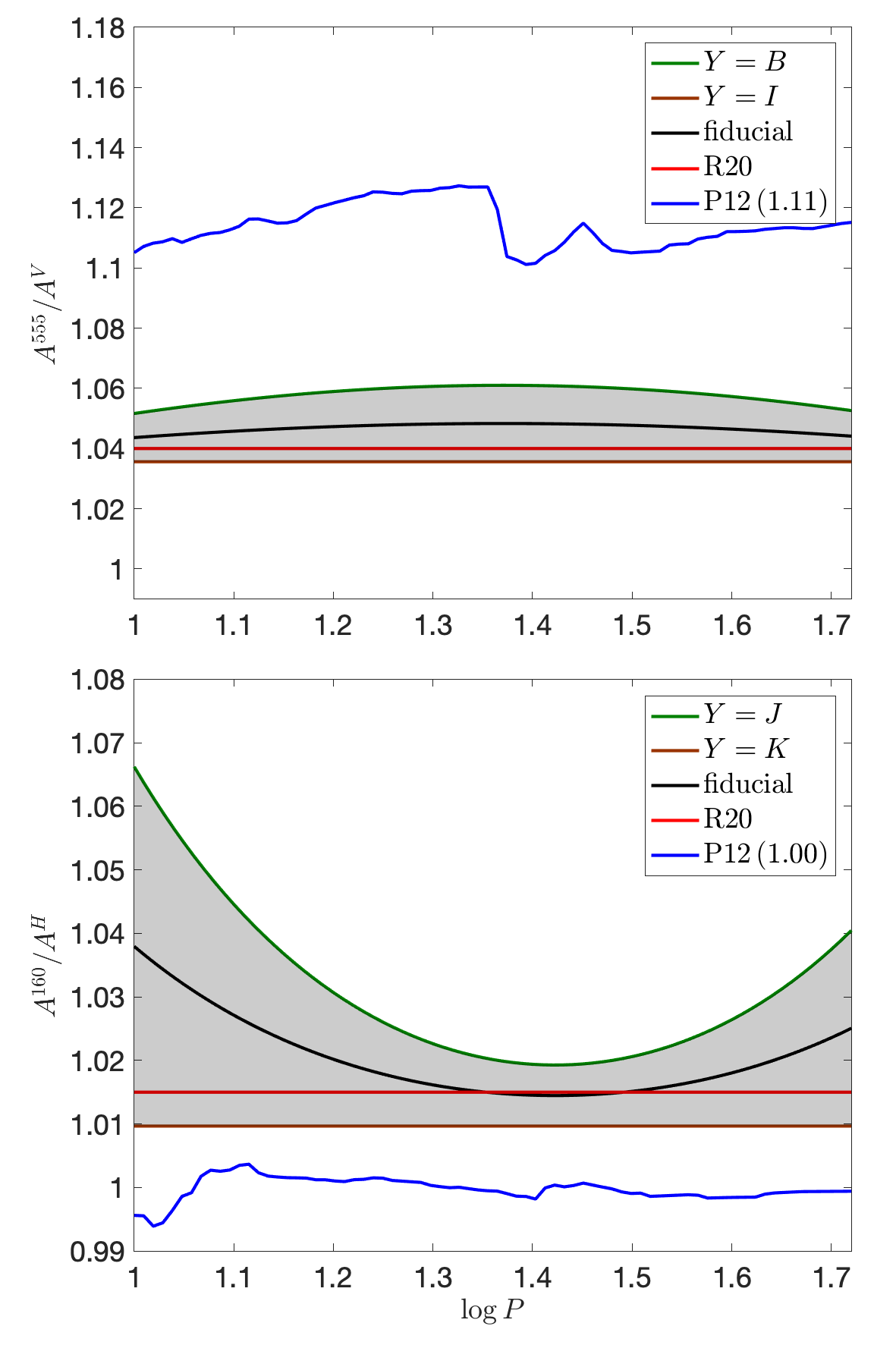}
	\caption{Top panel: $A^{555}/A^{V}$ as a function of the period. We can bound $A^{555}/A^{V}$ (dark region) between $(A^{555}/A^{V})_{B}\approx1.05-1.06$ from $Y=B$ (green line) and between $(A^{555}/A^{V})_{I}\approx1.035$ from $Y=I$ (brown line). The ratio $A^{555}/A^{V}=1.04$ used in R20 (red line), is within our bounded region. We interpolate between the two estimates with $A^{555}/A^{V}=f\left(A^{555}/A^{V}\right)_B+(1-f)\left(A^{555}/A^{V}\right)_I$, where $f=0.5$ is our fiducial value (black line) and the error is estimated with $f=0$ and $f=1$. The P12 templates (blue line) predict a value which is larger from our estimate by $\myapprox5\%$. This deviation could be related to less precise prediction of the P12 templates for \textit{HST} filters (see Section~\ref{sec:HST amplitude}). Bottom panel: $A^{160}/A^{H}$ as a function of the period. We can bound $A^{160}/A^{H}$ (dark region) between $(A^{160}/A^{H})_{J}\approx1.02-1.065$ from $Y=J$ (green line) and between $(A^{160}/A^{H})_{K}\approx1.01$ from $Y=K$ (brown line). The ratio $A^{160}/A^{H}=1.015$ used in R20, is within our bounded region. We interpolate between the two estimates with $A^{160}/A^{H}=f\left(A^{160}/A^{H}\right)_J+(1-f)\left(A^{160}/A^{H}\right)_K$, where $f=0.5$ is our fiducial value (black line) and the error is estimated with $f=0$ and $f=1$. The P12 templates (blue line) predict a value which is smaller from our estimate by $\myapprox1\%$.}
	\label{fig:HSTtoGround}
\end{figure}

The results of the interpolations (extrapolations) with 
\begin{equation}\label{eq:XYZ tran}
\left(\frac{A^{Z}}{A^{X}}\right)_{Y}=\frac{\left(\frac{A^{Y}}{A^{X}}-1\right)\left(\frac{\lambda_{X}}{\lambda_{Z}}-1\right)+\frac{\lambda_{X}}{\lambda_{Y}}-1}{\frac{\lambda_{X}}{\lambda_{Y}}-1}
\end{equation}
for $A^{555}/A^{V}$ are presented in the top panel of Figure~\ref{fig:HSTtoGround}. As can be seen, we can bound $A^{555}/A^{V}$ (dark region) between $(A^{555}/A^{V})_{B}\approx1.05-1.06$ from $Y=B$ (green line) and between $(A^{555}/A^{V})_{I}\approx1.035$ from $Y=I$ (brown line). The ratio $A^{555}/A^{V}=1.04$ used in R20, is within our bounded region. In what follows we interpolate between the two estimates with $A^{555}/A^{V}=f\left(A^{555}/A^{V}\right)_B+(1-f)\left(A^{555}/A^{V}\right)_I$, where $f=0.5$ is our fiducial value (black line) and the error is estimated with $f=0$ and $f=1$. The P12 templates (blue line) predict a value which is larger from our estimate by $\myapprox5\%$. This deviation could be related to less precise prediction of the P12 templates for \textit{HST} filters (see Section~\ref{sec:HST amplitude}).

As can be seen in the bottom panel of Figure~\ref{fig:HSTtoGround}, we can bound $A^{160}/A^{H}$ (dark region) between $(A^{160}/A^{H})_{J}\approx1.02-1.065$ from $Y=J$ (green line) and between $(A^{160}/A^{H})_{K}\approx1.01$ from $Y=K$ (brown line). The ratio $A^{160}/A^{H}=1.015$ used in R20, is within our bounded region. In what follows we interpolate between the two estimates with $A^{160}/A^{H}=f\left(A^{160}/A^{H}\right)_J+(1-f)\left(A^{160}/A^{H}\right)_K$, where $f=0.5$ is our fiducial value (black line) and the error is estimated with $f=0$ and $f=1$. The P12 templates (blue line) predict a value which is smaller from our estimate by $\myapprox1\%$. 

\begin{figure}
	\includegraphics[width=\columnwidth]{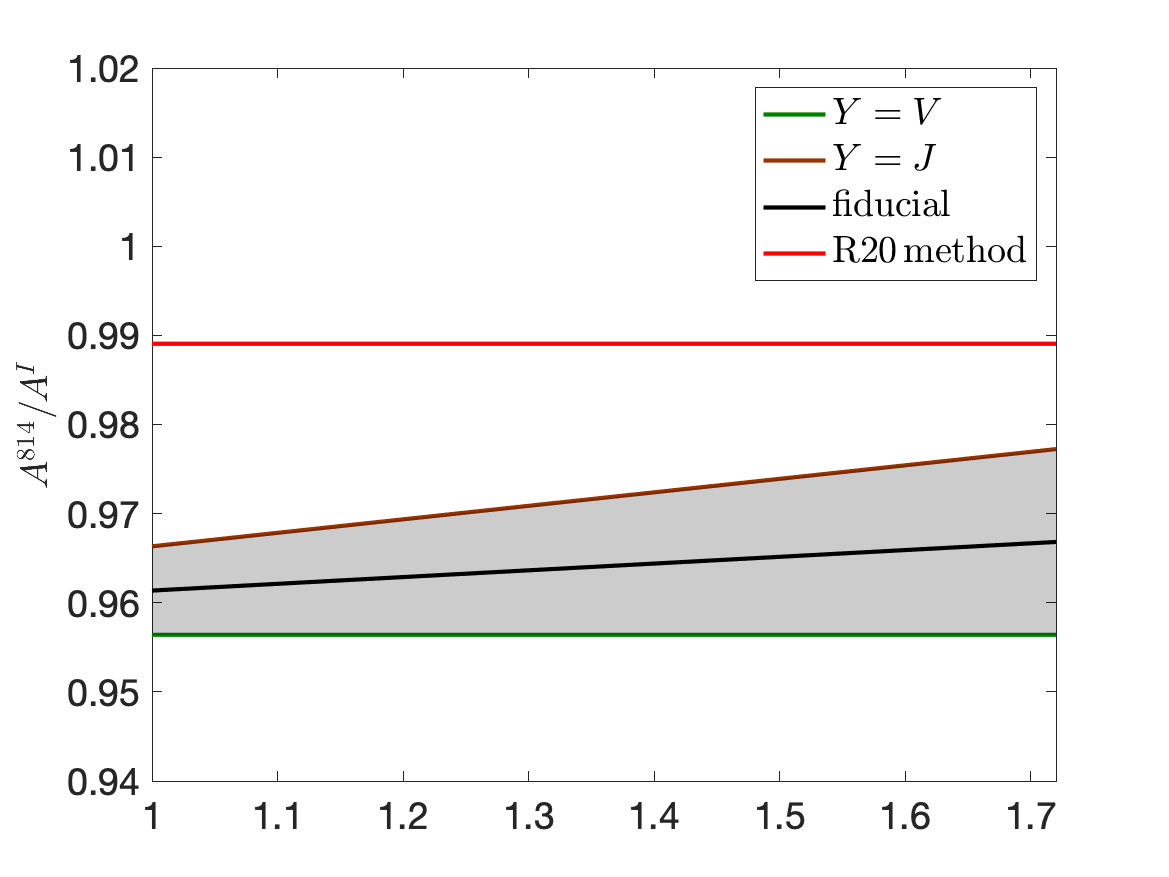}
	\caption{$A^{814}/A^{I}$ as a function of the period. We can bound $A^{814}/A^{I}$ (dark region) between $(A^{814}/A^{I})_{V}\approx0.955$ from $Y=B$ (green line) and between $(A^{555}/A^{V})_{I}\approx0.965-0.98$ from $Y=J$ (brown line). The ratio $A^{814}/A^{I}\approx0.99$, derived by using the R20 method with the relation $F814W$=$I+0.02-0.018(V-I)$ from R16, overpredicts our estimate by $\myapprox3\%$. This deviation could be related to the problems with the R20 method discussed in the text. We interpolate between the two estimates with $A^{814}/A^{I}=f\left(A^{814}/A^{I}\right)_V+(1-f)\left(A^{814}/A^{I}\right)_J$, with $f=0.5$ for the fiducial value (black line) and the error can be estimated with $f=0$ and $f=1$.}
	\label{fig:HSTtoGround814}
\end{figure}

We finally inspect the ratio $A^{160}/A^{I}$ (not required for our analysis) in Figure~\ref{fig:HSTtoGround814}. As can be seen in the figure, we can bound $A^{814}/A^{I}$ (dark region) between $(A^{814}/A^{I})_{V}\approx0.955$ from $Y=B$ (green line) and between $(A^{555}/A^{V})_{I}\approx0.965-0.98$ from $Y=J$ (brown line). The ratio $A^{814}/A^{I}\approx0.99$, derived by using the R20 method with the relation $F814W$=$I+0.02-0.018(V-I)$ from R16 (their equation 11), overpredicts our estimate by $\myapprox3\%$. This deviation could be related to the problems with the R20 method discussed above. One can interpolate between the two estimates with $A^{814}/A^{I}=f\left(A^{814}/A^{I}\right)_V+(1-f)\left(A^{814}/A^{I}\right)_J$, with $f=0.5$ for the fiducial value (black line) and the error can be estimated with $f=0$ and $f=1$. 

\section{Simulations of the process of measuring amplitude ratios, $A^{555}/A^{350}$, in a distant galaxy like NGC 5584}
\label{sec:simulation}

\begin{figure*}
	\includegraphics[width=\textwidth]{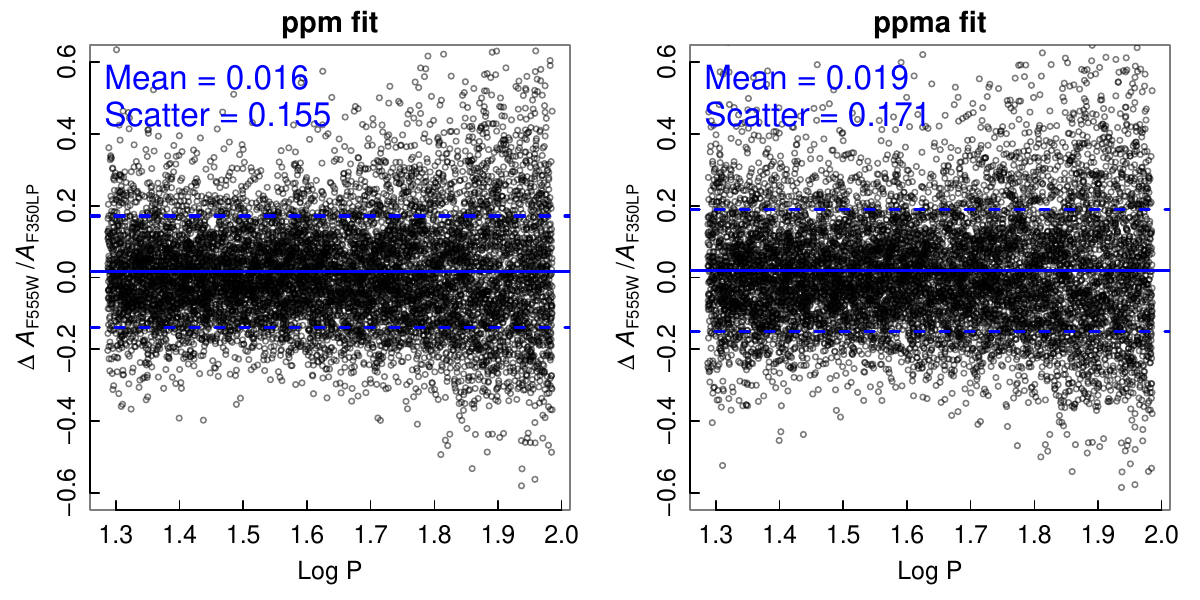}
	\includegraphics[width=\textwidth]{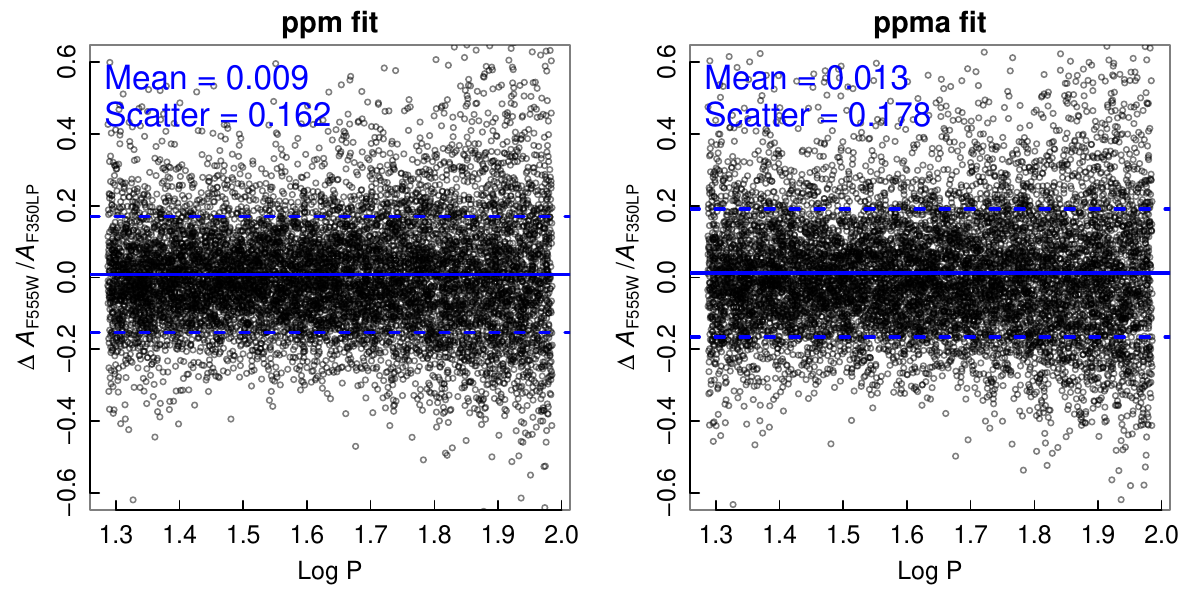}
	\caption{Fitted amplitude ratio, $A^{555}/A^{350}$, difference, where $\Delta=$output-input as a function of $\logp$ obtained in simulations of the process of measuring amplitude ratios in a distant galaxy like NGC 5584. PPM method (left panels): solve for the best-fitting period, phase and three mean magnitudes and once found optimize these fits for three amplitudes. PPMA method (right panels): optimize all $8$ parameters simultaneously. Upper panels: input amplitude was the same as the template. Lower panels: randomized amplitude parameter (with amplitude ratios from H16 used to scale the other bands). All methods show similar results. The PPMA method has slightly larger errors because all parameters are determined simultaneously. Neither method produces a bias in the period or mean magnitudes. The amplitudes are measured with a mean precision of $\mysim0.07-0.08$ per band (similar to what we found in the real data) with no significant bias to the precision of the test. There is a small bias in the fitted amplitude ratio of $\mysim0.015$, which is significant given the precision of the test with $10000$ fakes. The sense of this bias is a small overestimate of the amplitude ratios from measured data.}
	\label{fig:test}
\end{figure*}

In this appendix, we performe simulations of the process of measuring amplitude ratios, $A^{555}/A^{350}$, in a distant galaxy like NGC 5584, as was done in H16. We randomly selected a period from the observed range which defines the \citet{Yoachim2009} light curve template in $3$ bands, $F555W$($V$), $F814W$($I$) and $F350LP$($V$). Using the same light curve sampling and realistic noise as for NGC 5584, we produced noisy light curves and fit them with the templates. We did the fitting two ways. Method one (often used in past work) was to solve for the best-fitting period, phase and three mean magnitudes and once found optimize these fits for three amplitudes (PPM method). The second approach which is more computationally intensive is to optimize all $8$ parameters simultaneously (PPMA method). The results for recovering the amplitude ratio, shown in Figure~\ref{fig:test}, are quite similar for the two methods. The PPMA method has slightly larger errors because all parameters are determined simultaneously. Further, we performed this test two ways: 1) input amplitude was the same as the template and 2) a randomized amplitude parameter (with amplitude ratios from H16 used to scale the other bands). These two tests also yielded similar results. Neither produces a bias in the period or mean magnitudes. The amplitudes are measured with a mean precision of $\mysim0.07-0.08$ per band (similar to what we found in the real data) with no significant bias to the precision of the test. There is a small bias in the fitted amplitude ratio, $A^{555}/A^{350}$, where $\Delta=$output-input has a mean of $\mysim0.015$ (see Figure~\ref{fig:test}), which is significant given the precision of the test with $10000$ fakes. The sense of this bias is a small overestimate of the amplitude ratios from measured data. 

\section{The H16 amplitude distributions of $A^{555}$ and $A^{350}$ in different period bins}
\label{sec:histogram}

\begin{figure*}
	\includegraphics[width=1\textwidth]{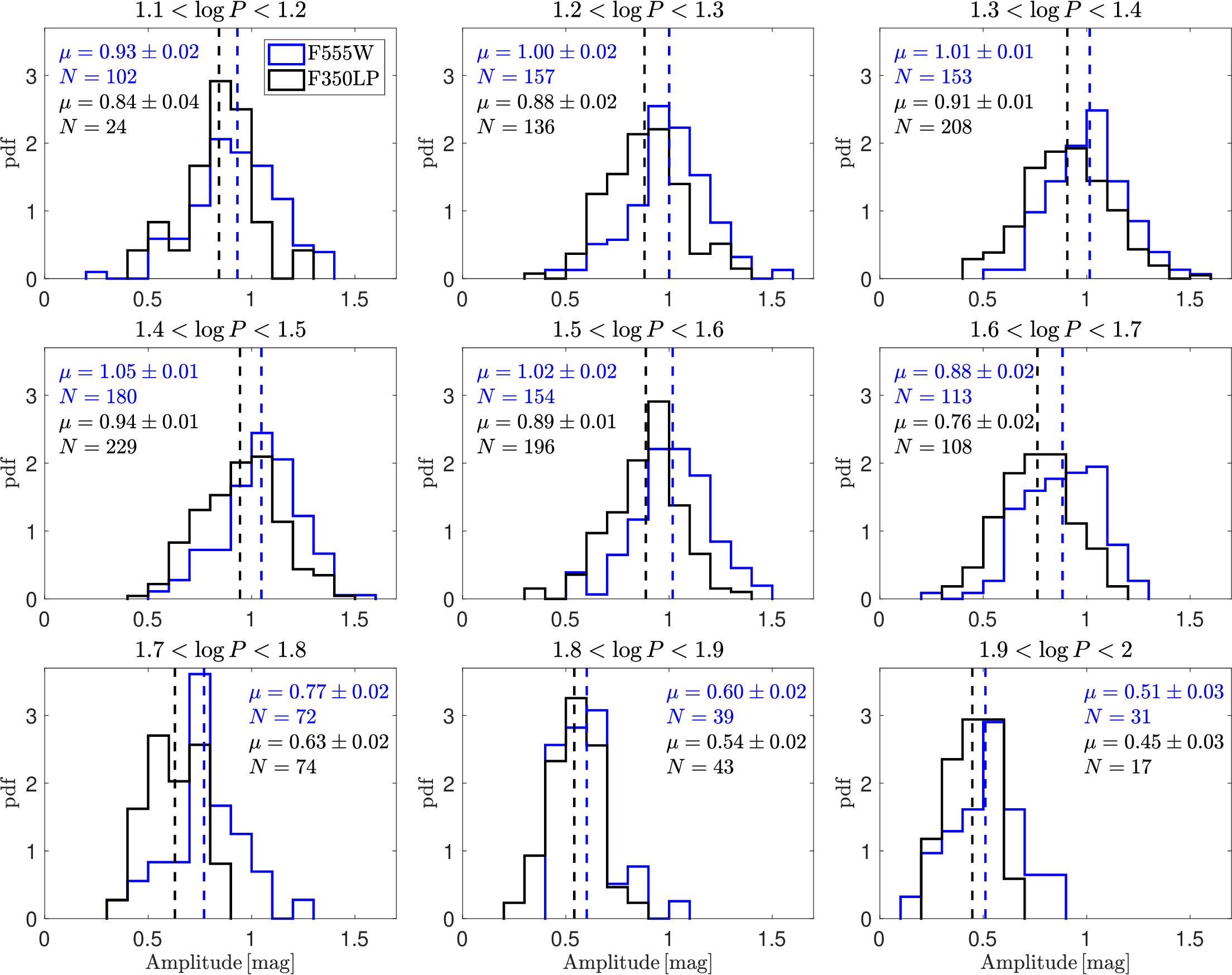}
	\caption{The H16 amplitude distributions of $A^{555}$ (blue) and $A^{350}$ (black) in each period bin presented in Figure~\ref{fig:350Vs555}. In each period bin the entire $A^{555}$ distribution is shifted from the $A^{350}$ distribution to higher amplitudes. The means of the distributions (dashed lines, corresponds to the red and magenta symbols with error bars in the middle panel of Figure~\ref{fig:350Vs555}) and the number of Cepheids in each period bins are indicated as well.}
	\label{fig:350Vs555_hist}
\end{figure*}

In this appendix, we supplement the claim in Section~\ref{sec:HST amplitude} that the means of the H16 amplitude distributions of $A^{555}$ are consistently larger than the means of $A^{350}$ by presenting the full H16 amplitude distributions of $A^{555}$ and $A^{350}$ in each period bin presented in Figure~\ref{fig:350Vs555}. The distributions are presented in Figure~\ref{fig:350Vs555_hist}. As can be seen in the figure, in each period bin the entire $A^{555}$ distribution is shifted from the $A^{350}$ distribution to higher amplitudes.

\end{appendix}
\bsp	
\label{lastpage}
\end{document}